\documentclass[useAMS,usenatbib]{mn2e}
\usepackage{graphicx}
\title[Electron impact excitation of N~IV]{Electron impact excitation of  N~IV: calculations with the DARC code and\, a comparison with ICFT results}
\author[K. M. Aggarwal,  F. P. Keenan and K. D. Lawson]{K.  M.  ~Aggarwal$^{1}$\thanks{E-mail:
 K.Aggarwal@qub.ac.uk (KMA);  F.Keenan@qub.ac.uk (FPK); Kerry.Lawson@ukaea.uk (KDL)}, F.  P.   ~Keenan$^{1}$ and K. D. Lawson$^{2}$   \\
$^{1}$Astrophysics Research Centre, School of Mathematics and Physics, Queen's University Belfast, Belfast BT7 1NN,  UK \\
$^{2}$CCFE, Culham Science Centre, Abingdon, OX14 3DB, UK}
\begin{document}

\date{Accepted 2016 June 3. Received 2016 June 2; in original form 2016 April 14}

\pagerange{\pageref{firstpage}--\pageref{lastpage}} \pubyear{2016}

\maketitle

\label{firstpage}

\begin{abstract}
There have been discussions in the recent literature regarding the accuracy of the available electron impact excitation rates (equivalently effective collision strengths $\Upsilon$) for transitions in Be-like ions.  In the present paper we demonstrate, once again, that earlier  results for $\Upsilon$ are indeed overestimated (by up to four orders of magnitude), for over 40\% of transitions and over a wide range of temperatures. To do this we have performed two sets of calculations for N~IV, with two different model sizes consisting of 166 and 238 fine-structure energy levels.   As in our previous work, for the determination of atomic structure the {\sc grasp} (General-purpose Relativistic Atomic Structure Package) is adopted and for the scattering calculations (the standard and parallelised versions of) the Dirac Atomic R-matrix Code ({\sc darc}) are employed. Calculations for collision strengths and effective collision strengths have been performed over a wide range of energy (up to 45~Ryd) and temperature (up to 2.0$\times$10$^6$~K), useful for applications in a variety of plasmas. Corresponding results for energy levels, lifetimes and A-values for all E1, E2, M1 and M2 transitions among 238 levels of N~IV are also reported. 

\end{abstract}


\section{Introduction}

Emission lines from Be-like ions have provided useful electron density and temperature  diagnostics for a variety of astrophysical plasmas. Many ions in this series, such as C~III,  N~IV, Ti~XIX and Fe~XXIII, are also important for the study of fusion plasmas. For N~IV \cite{vhc} have measured the 2s$^2$ $^1$S$_0$--2s2p $^1$P$^o_1$ line at 76.5 nm  in the Swarthmore Spheromak Experiment to diagnose  the plasma impurities.  Similarly, \cite{bra} have measured three impurity lines of N~IV ($\lambda$ 765, 923 and 1719 $\rm \AA$) in the NOVA-UNICAMP tokamak plasma. However, for plasma modelling  accurate atomic data are required, particularly for energy levels,  radiative rates (A-values), and excitation rates or equivalently the effective collision strengths ($\Upsilon$), which are obtained from the electron impact collision strengths ($\Omega$).  Given that, we have already reported such data for a number of Be-like ions, namely C~III,  Al~X, Cl~XIV, K~XVI, Ti~XIX and Ge~XXIX -- see \cite{belike, c3} and references therein. In this paper we focus our attention on N~IV.

Several emission lines of N~IV have been observed in the Sun \citep*{ddk}. In addition,  forbidden  lines, mostly belonging to the 2s$^2$ $^1$S -- 2s2p $^3$P$^o$ multiplet ($\lambda$$\lambda$ 1483,1486 $\rm \AA$), have been observed in the ultraviolet  spectra of low-density astrophysical plasmas, such as planetary nebulae and symbiotic stars -- see for example, \cite*{waf} and \cite{df}. These doublet lines are also detected in Ly$\alpha$ emitting galaxies (\cite{fos} and \cite{vanz}) and in low-mass and low-luminosity galaxies \citep{stark}.

An early analysis of solar emission lines of N~IV was undertaken by \cite{ddk}. In the absence of direct calculations of collisional data, they interpolated  $\Upsilon$ from the existing results for C~III and O~V. Subsequently, \cite{rams} calculated such data adopting the R-matrix code. These calculations  are in $LS$  coupling (Russell-Saunders or spin-orbit coupling) and include the 12 states of the 2s$^2$, 2s2p, 2p$^2$ and 2s3$\ell$ configurations. Subsequent results for  fine-structure transitions were obtained through an algebraic re-coupling scheme, and  stored in an  earlier version of the CHIANTI database at {\tt http://www.chiantidatabase.org/}.

Recently, \cite*{icft} have performed much larger calculations for a series of Be-like ions, including N~IV.  They have considered 238 fine-structure levels, belonging to the $n \le$ 7 configurations.  For the generation of wave functions, i.e. to determine energy levels and A-values, they adopted the {\em AutoStructure} (AS) code of \cite{as}  and  for  the subsequent calculations of $\Omega$ and $\Upsilon$,  the R-matrix code of \cite*{rm2}. However, they also  primarily obtained  $\Omega$ in  $LS$ coupling, and the corresponding results for  fine-structure transitions  were determined  through their intermediate coupling frame transformation (ICFT) method, similar to the one adopted by \cite{rams}.  The AS code does not include higher-order relativistic effects, which are important for heavier systems (such as Ge~XXIX),  but not for a  comparatively light ion, such as N~IV. Therefore, their calculations should represent a significant extension and improvement over the earlier results of \cite{rams}. However, our work on a number of Be-like ions (Al~X, Cl~XIV, K~XVI, Ti~XIX  and Ge~XXIX) indicated that the data of \cite{icft}  were highly overestimated for a significant number of transitions, and over a wide range of electron temperatures \citep{belike}. These Be-like ions are comparatively highly ionised, but similar discrepancies have also been noted for transitions in C~III \citep{c3} as well as for Al-like Fe XIV \citep{fe14}, and most recently for Ar-like Fe IX \citep{tz}.

No two independent atomic data calculations are ever exactly the same, but there are two  major differences between our work and that of \cite{icft}, namely the methodology and the size. For the scattering calculations we have adopted the fully relativistic  Dirac atomic R-matrix code  ({\sc darc}), in comparison to their semi-relativistic R-matrix method through the ICFT approach. However, in principle both approaches should provide comparable results for a majority of transitions and over a wide range of temperature, as has already been observed in several cases -- see for example the work of \cite{bb} on Fe~III and references therein.  Therefore, the discrepancies noted in the values of $\Upsilon$ for a range of Be-like ions are perhaps not due to the methodologies but their implementation, as already discussed in detail  by us \citep{belike}.  Since such large discrepancies are worrying and need to be addressed so that data can be confidently applied to plasma modelling, \cite{al10} made an extensive analysis of these discrepancies taking Al~X as an example. They rather concluded that the differences in the calculations of $\Upsilon$ lie in the corresponding differences in the determination of {\em atomic structure},\, and not in the implementation of the scattering methods as we suggested. Since they could perform much larger calculations than us (and indeed others such as \cite{tz}), they not only defended their work but also concluded their results to be more accurate. In general, it is undoubtedly true that a larger calculation should be superior (i.e. comparatively more accurate), because of  the inclusion of resonances arising from the higher-lying levels of the additional  configurations. However, these should not increase  values of $\Upsilon$ by orders of magnitude, and not for a significant number of transitions and over an entire range of temperatures. Unfortunately, until now it was not possible for us to match the size of the \cite{icft} calculations, because of the limitations of the computational resources available to us. However, one of our colleagues  (Dr. Connor Ballance) has now implemented the parallelised version of the DARC code and therefore we are able to perform as large a calculation as  \cite{icft,al10}. We do this for an important Be-like ion, i.e.  N~IV, to make direct comparisons with their work.

\begin{table*}
 \centering
 \caption{Energy levels (in Ryd) of N~IV and their lifetimes (s).  $a{\pm}b \equiv a{\times}$10$^{{\pm}b}$.}
\begin{tabular}{rllrrrrrrll} \hline
Index  & Configuration       & Level &  NIST      &   GRASP1  &  GRASP2 & AS         & GRASP1($\tau$, s)    & GRASP2($\tau$, s) \\
\hline
    1 &    2s$^2$  &  $^1$S$  _0$   & 0.00000   & 0.00000   & 0.00000 & 0.00000  & ..........   &  ......... \\
    2 &    2s2p    &  $^3$P$^o_0$   & 0.61246   & 0.61845   & 0.61794 & 0.62331  & ..........   & .......... \\
    3 &    2s2p    &  $^3$P$^o_1$   & 0.61303   & 0.61898   & 0.61848 & 0.62408  & 2.471$-$03   & 2.423$-$03 \\
    4 &    2s2p    &  $^3$P$^o_2$   & 0.61434   & 0.62024   & 0.61974 & 0.62563  & 8.439$+$01   & 8.447$+$01 \\
    5 &    2s2p    &  $^1$P$^o_1$   & 1.19097   & 1.26289   & 1.25872 & 1.26008  & 3.670$-$10   & 3.689$-$10 \\
    6 &    2p$^2$  &  $^3$P$  _0$   & 1.59960   & 1.62677   & 1.62620 & 1.63781  & 5.290$-$10   & 5.290$-$10 \\
    7 &    2p$^2$  &  $^3$P$  _1$   & 1.60026   & 1.62740   & 1.62683 & 1.63857  & 5.285$-$10   & 5.285$-$10 \\
    8 &    2p$^2$  &  $^3$P$  _2$   & 1.60140   & 1.62845   & 1.62788 & 1.64009  & 5.277$-$10   & 5.277$-$10 \\
    9 &    2p$^2$  &  $^1$D$  _2$   & 1.72122   & 1.78923   & 1.78560 & 1.79966  & 4.304$-$09   & 4.315$-$09 \\
   10 &    2p$^2$  &  $^1$S$  _0$   & 2.14484   & 2.26067   & 2.25835 & 2.26451  & 2.796$-$10   & 2.797$-$10 \\
    ... & \\
    ... & \\
    ... & \\
\hline  
\end{tabular}

\begin{flushleft}
{\small
NIST: {\tt http://www.nist.gov/pml/data/asd.cfm} \\
GRASP1: Energies from the {\sc grasp} code for 166 level calculations \\
GRASP2: Energies from the {\sc grasp} code for 238 level calculations \\
AS: Energies from the AS calculations \citep{icft} for 238 levels \\
}
\end{flushleft}
\end{table*}

\setcounter{table}{1}                                                                                                                                           
\begin{table*}                                                                                                                                                  
\caption{Transition wavelengths ($\lambda_{ij}$ in $\AA$), radiative rates (A$_{ji}$ in s$^{-1}$), oscillator strengths (f$_{ij}$, dimensionless), and line  
strengths (S, in atomic units) for electric dipole (E1), and A$_{ji}$ for E2, M1 and M2 transitions in N~IV. $a{\pm}b \equiv a{\times}$10$^{{\pm}b}$. See Table 1 for level indices. Complete table is available online as Supporting Information.}     
\begin{tabular}{rrrrrrrrr}                                                                                                                                      
\hline                                                                                                                                                                                                                                                                                                               
$i$ & $j$ & $\lambda_{ij}$ & A$^{{\rm E1}}_{ji}$  & f$^{{\rm E1}}_{ij}$ & S$^{{\rm E1}}$ & A$^{{\rm E2}}_{ji}$  & A$^{{\rm M1}}_{ji}$ & A$^{{\rm M2}}_{ji}$ \\  
\hline                                                                                                                                                   
    1 &    3 &  1.473$+$03 &  4.128$+$02 &  4.030$-$07 &  1.955$-$06 &  0.000$+$00 &  0.000$+$00 &  0.000$+$00 \\       
    1 &    4 &  1.470$+$03 &  0.000$+$00 &  0.000$+$00 &  0.000$+$00 &  0.000$+$00 &  0.000$+$00 &  1.180$-$02 \\       
    1 &    5 &  7.240$+$02 &  2.711$+$09 &  6.391$-$01 &  1.523$+$00 &  0.000$+$00 &  0.000$+$00 &  0.000$+$00 \\       
    1 &    7 &  5.602$+$02 &  0.000$+$00 &  0.000$+$00 &  0.000$+$00 &  0.000$+$00 &  6.996$-$03 &  0.000$+$00 \\       
    1 &    8 &  5.598$+$02 &  0.000$+$00 &  0.000$+$00 &  0.000$+$00 &  7.809$-$02 &  0.000$+$00 &  0.000$+$00 \\       
    1 &    9 &  5.103$+$02 &  0.000$+$00 &  0.000$+$00 &  0.000$+$00 &  3.787$+$03 &  0.000$+$00 &  0.000$+$00 \\  
    ... & \\
    ... & \\
    ... & \\
\hline                                                                                                                                                          
\end{tabular}                                                                                                                                                   
\end{table*}                                           

\section[]{Energy levels}

As in our earlier work on other Be-like ions, we have  employed the fully relativistic {\sc grasp} (General-purpose Relativistic Atomic Structure  Package)  to  determine the atomic structure, i.e. to calculate energy levels and A-values. Measurements of energy levels for N~IV have been compiled and critically evaluated by the  NIST (National Institute of Standards and Technology) team \citep{nist} and are available at their  website {\tt http://www.nist.gov/pml/data/asd.cfm}. However, these energies are restricted to mostly low-lying levels and are missing for many of the $n \ge$ 4 configurations -- see Table 1. Theoretical energies have been determined by   several workers --  see for example, \cite{gu} and references therein -- but these are also restricted to a few lower levels, mostly up to $n$ = 3. However, as stated earlier, \cite{icft} have determined energies for 238 levels belonging to the $n \le$ 7 configurations.  In our work, we have performed two sets of calculations, i.e. GRASP1:  which includes 166 levels  of 27  configurations, namely (1s$^2$) 2$\ell$2$\ell'$, 2$\ell$3$\ell'$,  2$\ell$4$\ell'$  and 2$\ell$5$\ell'$.  These calculations are similar to those for  C~III \citep{c3}, but larger than for other Be-like ions we have investigated  \citep{belike}, which were confined to the lowest 98 levels. For the other (GRASP2) calculation we include the same 238 levels as by \cite{icft}, the additional 72 levels belonging to  (1s$^2$) 2s6s/p/d, 2p6s/p/d, 2s7s/p/d, and 2p7s/p/d, i.e. 39 configurations in total. Both calculations have been performed in an `extended average level' approximation and include contributions from the Breit and QED (quantum electrodynamic) effects. These energies are listed in Table~1 along with those of NIST and \cite{icft}.

The GRASP1 and GRASP2 energies are nearly the same, in both magnitude and ordering. Similarly, there is a general agreement (within 0.02 Ryd or 1\% for all levels) between our  GRASP  and the earlier AS  energies of \cite{icft}, and the orderings are also nearly the same for most levels with only a few exceptions -- see for example, levels 94--95, 98--101 and 149--151.  However, differences with the  NIST energies are significant (up to 6\%), particularly for the lowest 10 levels of the 2s2p and 2p$^2$ configurations -- see Fig. 1. Fortunately, discrepancies for the remaining  levels are  smaller than 1\%.  We also note that level 107 (2p4p $^3$S$_1$) is an exception, because its placing in the NIST listings is anomalous with results to that from  GRASP and AS, and theoretical energies for this level are lower by $\sim$0.1 Ryd. In the absence of any other calculation it is difficult to resolve its position, and results for C~III do not help  because the level orderings of the two ions are very different -- see table 1 of \cite{c3}.  However, all four $J$=1 levels of the 2p4p configuration, i.e. 94 ($^1$P$_1$), 96 ($^3$D$_1$), 107 ($^3$S$_1$) and 109 ($^3$P$_1$), are highly mixed, and interchanging their positions will not resolve the discrepancy in the energies, as none has a value closer to that of NIST. Additionally, the level mixing is strong only in $jj$ coupling and there is no ambiguity in $LSJ$ coupling. Finally,  we observe better  agreement between theoretical and experimental energies for the levels of N~IV  than for C~III \citep{c3},   but scope remains for improvement. An  inclusion of {\em pseudo} orbitals in the generation of wave functions may improve the accuracy of energy levels, but it will give rise to pseudo resonances in the subsequent scattering calculations for $\Omega$. Therefore, both ourselves and \cite{icft} have avoided this approach because the focus is on electron impact excitation.

\section{Radiative rates}

Generally, A-values for electric dipole (E1) transitions alone are not sufficient  for plasma modelling applications, even though they have larger magnitudes in comparison to other types, namely electric quadrupole (E2), magnetic dipole (M1) and  magnetic quadrupole (M2). Hence for completeness and also  for the accurate determination of lifetimes (see section 4) we have calculated  A-values  for all four types of transitions. Furthermore, although A-values are often directly employed in plasma modelling calculations, it is the  absorption oscillator strength (f$_{ij}$) which gives a general idea about the strength of a transition. However, the two parameters, for all types of  transition $i \to j$, are related by the following expression:

\begin{equation}
f_{ij} = \frac{mc}{8{\pi}^2{e^2}}{\lambda^2_{ji}} \frac{{\omega}_j}{{\omega}_i} A_{ji}
 = 1.49 \times 10^{-16} \lambda^2_{ji}  \frac{{\omega}_j}{{\omega}_i}  A_{ji} 
\end{equation}
where $m$ and $e$ are the electron mass and charge, respectively, $c$ the velocity of light,  $\lambda_{ji}$  the transition energy/wavelength in $\rm \AA$, and $\omega_i$ and $\omega_j$  the statistical weights of the lower ($i$) and upper ($j$) levels, respectively. Similarly, these two parameters are related to  the line strength S (in atomic unit, 1 a.u. = 6.460$\times$10$^{-36}$ cm$^2$ esu$^2$) by the following expression for E1 transitions:

\begin{equation}
A_{ji} = \frac{2.0261\times{10^{18}}}{{{\omega}_j}\lambda^3_{ji}} S^{{\rm E1}} \hspace*{0.5 cm} {\rm and} \hspace*{0.5 cm} 
f_{ij} = \frac{303.75}{\lambda_{ji}\omega_i} S^{{\rm E1}}. \\
\end{equation}
Similar equations for other types of transition may be found in \cite{tixix}.

As for energy levels, we have also calculated A-values from both the GRASP1 and GRASP2 models. In Table~2 we list our calculated energies/wavelengths ($\lambda$, in $\rm \AA$), radiative rates (A$_{ji}$, in s$^{-1}$), oscillator strengths (f$_{ij}$, dimensionless), and line strengths (S, in a.u.)  for all 8212 E1 transitions among the 238 levels of  N~IV. These results are  in the length  form because of their comparatively higher  accuracy.  The  indices used  to represent the lower and upper levels of a transition are defined in Table~1. Similarly, there  are 10~301 E2, 8136 M1  and 10~353 M2 transitions  among the same 238 levels, i.e. the GRASP2 model. Corresponding results from the GRASP1 model among 166 levels can be obtained from the first author (KMA) on request. Additionally,  only the A-values are listed in Table~2 for the E2, M1 and M2 transitions, and the corresponding results for f- values can be easily obtained through Eq. (1). 

A general criterion to assess  the accuracy of A-values is to look at the ratio (R) of their velocity and length forms. If R is close to unity then the A- (or f-) value is considered to be accurate, although it is only a desirable criterion, not a necessary nor indeed sufficient one. For most (strong) E1 transitions with f $\ge$ 0.01, the two forms normally give R$\sim$1 and their magnitudes do not significantly vary with differing amount of CI (configuration interaction) and/or methods. Among comparatively strong E1 transitions in N~IV,  for about a third  R differs   from unity by  more than $\pm$20\%. For most such transitions R is within a factor of two, but for a few  it has values up to an order of magnitude. Examples of transitions for which R is large are: 5--217 (f = 0.021), 9--81 (f =  0.012) and 20--185 (f = 0.014), i.e. all such transitions are invariably weak. For a few  very weak transitions (f $\sim$ 10$^{-5}$ or less) the two forms of f- values differ by up to several orders of magnitude, as also noted for transitions of C~III \citep{c3} and other Be-like ions. Nevertheless, such transitions with very small f-values are unlikely to significantly affect the modelling of plasmas.

Most of the A-values available in the literature for N~IV involve levels of the $n \le$ 3 configurations -- see for example, \cite{uiss} and \cite*{uis}. However, as for energy levels and collisional data, \cite{icft} have reported  results for a larger number of E1 transitions. For most transitions there is satisfactory  agreement between the two calculations, but for a few weak(er) ones  there are discrepancies  of over 50\%. Some examples are shown  in Table~3, in which results from both the GRASP1 and GRASP2 models are listed. Such discrepancies  for weak(er) transitions between any two calculations are quite common (see for example \cite{c3} for transitions of C~III) and often arise due to the different levels of CI as well as  the method adopted -- see particularly the weak transitions 2--37, 3--37 and 4--37.  Differences in CI may result in cancellation or addition of different matrix elements and hence affecting the A-(f-) values, particularly for the weak (inter-combination) transitions. However, both the GRASP and AS calculations include the same CI and therefore, the differences noted for transitions in Table~3 are mainly due to the methodology adopted. The A-values for a few M1 transitions are also available in the literature,  by \cite{glass} and \cite{uis}, and in Table~4 we compare our A-values (from GRASP1 and GRASP2) for the transitions in common. As for the weak E1 transitions, for the M1 ones  there are  also large discrepancies for a few, although the GRASP1 and GRASP2 A-values are very similar. In general, there is a closer agreement between our calculations and those of \cite{glass}, and the corresponding results of \cite{uis} differ by up to an order of magnitude (see for example the 7--9 transition).

\begin{figure*}
\includegraphics[angle=-90,width=0.9\textwidth]{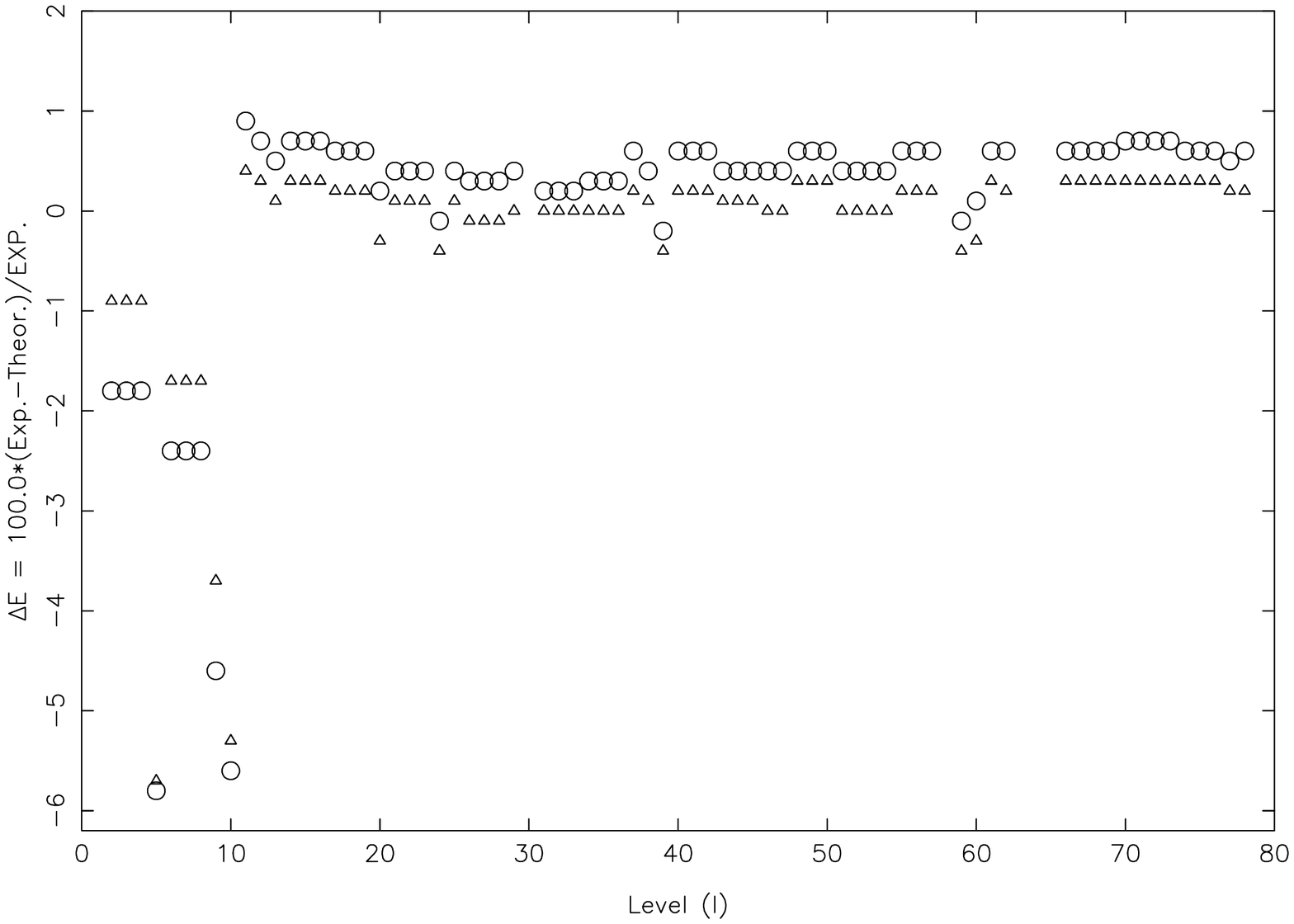}
 \vspace{-1.5cm}
 \caption{Percentage differences between experimental and theoretical energy levels of N~IV. Triangles: present GRASP2 calculations  and circles: calculations of  Fern{\'a}ndez-Menchero et al. (2014) with AS.}
\end{figure*}

\setcounter{table}{2}
\begin{table*}
\caption{Comparison of A-values for a few E1 transitions of N~IV.  $a{\pm}b \equiv a{\times}$10$^{{\pm}b}$.   See Table 1 for level indices.}
\begin{tabular}{rrrrrrrrrr} \hline
I & J& f (GRASP1) & A (GRASP1) & f (GRASP2) & A (GRASP2) &  A (AS) & R1 & R2  \\
\hline
    1   &  43  &  1.699$-$05  &  9.638$+$05  &  1.8916$-$05  &  1.0730$+$06  &  1.930$+$06  &	2.0 &	   1.8 \\
    2   &  37  &  1.753$-$06  &  7.183$+$04  &  2.3612$-$08  &  9.6752$+$02  &  2.720$+$06  &  37.9 &	2811.3 \\
    3   &  12  &  1.401$-$08  &  2.868$+$03  &  1.2904$-$08  &  2.6423$+$03  &  1.680$+$03  &	1.7 &	   1.6 \\
    3   &  20  &  2.740$-$07  &  1.441$+$04  &  2.7037$-$07  &  1.4225$+$04  &  2.480$+$04  &	1.7 &	   1.7 \\
    3   &  30  &  7.638$-$05  &  2.770$+$07  &  8.0979$-$05  &  2.9377$+$07  &  1.430$+$07  &	1.9 &	   2.1 \\
    3   &  37  &  1.425$-$07  &  1.750$+$04  &  2.3799$-$06  &  2.9248$+$05  &  4.000$+$06  & 228.6 &	  13.7 \\
    3   &  62  &  3.799$-$09  &  1.735$+$03  &  4.3647$-$10  &  1.9925$+$02  &  7.850$+$02  &	2.2 &	   3.9 \\
    3   &  77  &  1.012$-$07  &  9.545$+$03  &  5.2196$-$08  &  4.9224$+$03  &  4.370$+$03  &	2.2 &	   1.1 \\
    4   &  37  &  1.489$-$05  &  3.048$+$06  &  2.5151$-$05  &  5.1482$+$06  &  1.110$+$05  &  27.5 &	  46.4 \\
    4   &  70  &  1.151$-$12  &  1.287$-$01  &  9.8121$-$13  &  1.0976$-$01  &  3.960$-$01  &	3.1 &	   3.6 \\
    5   &   6  &  1.836$-$07  &  5.859$+$02  &  1.8553$-$07  &  6.0372$+$02  &  1.010$+$03  &	1.7 &	   1.7 \\
    5   &   7  &  3.061$-$08  &  3.267$+$01  &  3.1353$-$08  &  3.4126$+$01  &  5.560$+$01  &	1.7 &	   1.6 \\
    5   &   8  &  2.819$-$06  &  1.816$+$03  &  2.9598$-$06  &  1.9440$+$03  &  2.820$+$03  &	1.6 &	   1.5 \\
    5   &  11  &  1.590$-$07  &  5.976$+$03  &  1.6108$-$07  &  6.0762$+$03  &  9.430$+$03  &	1.6 &	   1.6 \\
    5   &  18  &  3.059$-$07  &  9.634$+$03  &  3.1387$-$07  &  9.9192$+$03  &  2.030$+$04  &	2.1 &	   2.0 \\
    5   &  31  &  1.932$-$05  &  4.890$+$06  &  1.6592$-$05  &  4.2118$+$06  &  2.860$+$06  &	1.7 &	   1.5 \\
    5   &  48  &  4.877$-$08  &  4.489$+$03  &  4.9811$-$08  &  4.5969$+$03  &  6.980$+$03  &	1.6 &	   1.5 \\
    5   &  49  &  4.652$-$09  &  2.569$+$02  &  5.2655$-$09  &  2.9157$+$02  &  1.180$+$03  &	4.6 &	   4.0 \\
    5   &  61  &  3.258$-$09  &  3.558$+$02  &  3.7610$-$09  &  4.1155$+$02  &  2.760$+$03  &	7.8 &	   6.7 \\
    5   &  67  &  1.938$-$08  &  2.200$+$03  &  1.9257$-$08  &  2.1914$+$03  &  4.250$+$03  &	1.9 &	   1.9 \\
    5   &  68  &  6.116$-$09  &  4.166$+$02  &  8.6323$-$09  &  5.8942$+$02  &  2.890$+$03  &	6.9 &	   4.9 \\
\hline  
\end{tabular}

\begin{flushleft}
{\small
GRASP1: Present 166 level calculations with the {\sc grasp} code \\
GRASP2: Present 238 level calculations with the {\sc grasp} code \\
AS: Calculations of ICFT  \citep{icft} with the {\sc as} code \\
R1: ratio of  GRASP1 and AS A-values, the larger of the two is always in the numerator \\
R2: ratio of  GRASP2 and AS A-values, the larger of the two is always in the numerator \\
}
\end{flushleft}
\end{table*}

\setcounter{table}{3}
\begin{table*}
\caption{Comparison of A-values for a few M1 transitions of N~IV.  $a{\pm}b \equiv a{\times}$10$^{{\pm}b}$.  See Table 1 for level indices.}
\begin{tabular}{rrcccc} \hline
I & J &  GRASP1 & GRASP2 &  \cite{glass}   &   \cite{uis}  \\ 
\hline
1  &   7  &  6.967$-$3 & 6.996$-$3  & 4.627$-$3   & 5.82$-$3   \\  
2  &   3  &  3.713$-$6 & 3.734$-$6  & 4.283$-$6   & 4.74$-$6   \\  
2  &   5  &  1.255$-$2 & 1.249$-$2  & 1.315$-$2   & 6.36$-$3   \\
3  &   4  &  3.535$-$5 & 3.545$-$5  & 3.762$-$5   & 4.11$-$5   \\
3  &   5  &  1.052$-$2 & 1.046$-$2  & 3.720$-$2   & 5.42$-$3   \\
4  &   5  &  1.637$-$2 & 1.630$-$2  & 1.629$-$2   & 8.42$-$3   \\
6  &   7  &  6.143$-$6 & 6.169$-$6  & 6.855$-$6   & 7.29$-$6   \\
7  &   8  &  2.041$-$5 & 2.057$-$5  & 2.440$-$5   & 2.81$-$5   \\
7  &   9  &  2.325$-$3 & 2.286$-$3  & 2.055$-$3   & 3.33$-$4   \\
7  &  10  &  1.318$-$1 & 1.319$-$1  & 1.572$-$1   & 5.51$-$2   \\
8  &   9  &  7.064$-$3 & 6.939$-$3  & 6.586$-$3   & 1.01$-$3   \\
\hline  
\end{tabular}

\begin{flushleft}
{\small
GRASP1: Present 166 level calculations with the {\sc grasp} code \\
GRASP2: Present 238 level calculations with the {\sc grasp} code \\
}
\end{flushleft}
\end{table*}

\setcounter{table}{4}
\begin{table*}
\caption{Comparison of lifetimes ($\tau$, ns) for a few levels of N~IV. } 
\begin{tabular}{lllllrrl} \hline
Configuration/Level  &  GRASP1 & GRASP2   & \cite{nl} \\ 
\hline
 2s2p $^1$P$^o$   & 0.367  & 0.369  &	0.44--0.53   \\
 2p$^2$ $^3$P	  & 0.529  & 0.529  &	0.60--0.70   \\
 2p$^2$ $^1$D	  & 4.304  & 4.315  &	3.10--4.73   \\
 2p$^2$ $^1$S	  & 0.280  & 0.280  &	0.34	     \\
 2s3s $^3$S	  & 0.109  & 0.109  &	0.13	     \\
 2s3s $^1$S	  & 0.316  & 0.326  &	0.38--0.40   \\
 2s3p $^3$P$^o$   & 7.845  & 7.898  &   7.3--11.5    \\
 2s3d $^3$D	  & 0.033  & 0.033  &	0.033--0.043 \\
 2s3d $^1$D	  & 0.053  & 0.053  &	0.050--0.14  \\
 2p3s $^3$P$^o$   & 0.135  & 0.136  &	0.15--0.30   \\
 2p3s $^1$P$^o$   & 0.112  & 0.113  &	0.13--0.30   \\
 2p3p $^1$P	  & 0.111  & 0.115  &	0.10--0.12   \\
 2p3p $^3$P	  & 0.183  & 0.183  &	0.18--7.83   \\
 2p3p $^1$D	  & 0.103  & 0.103  &	0.082--0.11  \\
 2p3p $^3$D	  & 0.257  & 0.261  &	0.22--0.355  \\
 2p3d $^3$D$^o$   & 0.027  & 0.028  &	0.031--0.23  \\
 2p3d $^3$P$^o$   & 0.064  & 0.064  &	0.078--0.62  \\
 2p3d $^1$F$^o$   & 0.044  & 0.045  &	0.067	     \\
 2s4s $^3$S	  & 2.991  & 2.960  &   0.12	     \\
 2s4p $^1$P$^o$   & 0.140  & 0.150  &	0.16--0.55   \\
 2s4p $^3$P$^o$   & 0.203  & 0.200  &	0.17	     \\
 2s4d $^3$D	  & 0.087  & 0.087  &	0.093--0.17  \\
 2s4d $^1$D	  & 0.121  & 0.128  &	0.12--0.9    \\
 2s4f $^3$F$^o$	  & 0.238  & 0.238  &	0.294--0.35  \\
 2s4f $^1$F$^o$	  & 0.058  & 0.058  &	0.075	     \\
 2s5s $^3$S	  & 0.326  & 0.359  &	0.37	     \\
 2s5p $^1$P$^o$   & 0.132  & 0.204  &	0.32	     \\
 2s5f $^3$F$^o$   & 0.444  & 0.438  &	0.43--2.4    \\
 2s5f $^1$F$^o$   & 0.337  & 0.384  &	0.48	     \\
 2s5g $^3$G	  & 0.918  & 0.918  &	0.82--1.22   \\
 2s5g $^1$G	  & 0.922  & 0.922  &	1.11--1.35   \\
\hline  
\end{tabular}

\begin{flushleft}
{\small
GRASP1: Present 166 level calculations with the {\sc grasp} code \\
GRASP2: Present 238 level calculations with the {\sc grasp} code \\
}
\end{flushleft}
\end{table*}

\setcounter{table}{5}
\begin{table*}
\caption{Comparison of lifetimes ($\tau$, s)  for the lowest 20 levels of N~IV.  $a{\pm}b \equiv a{\times}$10$^{{\pm}b}$. }
\begin{tabular}{rllrllrrl} \hline
Index  & Configuration       & Level & GRASP1 & GRASP2       &    \cite{tff}         \\  
\hline
    1  &    2s$^2$  &  $^1$S$  _0$   &  .....	    &   .........  &  .....	  \\ 
    2  &    2s2p    &  $^3$P$^o_0$   &  .....	    &  ..........  &  .....	  \\ 
    3  &    2s2p    &  $^3$P$^o_1$   &  2.471$-$03  &  2.423$-$03  &  1.726$-$03  \\
    4  &    2s2p    &  $^3$P$^o_2$   &  8.439$+$01  &  8.447$+$01  &  8.606$+$01  \\
    5  &    2s2p    &  $^1$P$^o_1$   &  3.670$-$10  &  3.689$-$10  &  4.306$-$10  \\
    6  &    2p$^2$  &  $^3$P$  _0$   &  5.290$-$10  &  5.290$-$10  &  5.606$-$10  \\
    7  &    2p$^2$  &  $^3$P$  _1$   &  5.285$-$10  &  5.285$-$10  &  5.660$-$10  \\
    8  &    2p$^2$  &  $^3$P$  _2$   &  5.277$-$10  &  5.277$-$10  &  5.651$-$10  \\
    9  &    2p$^2$  &  $^1$D$  _2$   &  4.304$-$09  &  4.315$-$09  &  4.266$-$09  \\
   10  &    2p$^2$  &  $^1$S$  _0$   &  2.796$-$10  &  2.797$-$10  &  3.399$-$10  \\
   11  &    2s3s    &  $^3$S$  _1$   &  1.085$-$10  &  1.090$-$10  &  1.108$-$10  \\
   12  &    2s3s    &  $^1$S$  _0$   &  3.158$-$10  &  3.261$-$10  &  3.826$-$10  \\
   13  &    2s3p    &  $^3$P$^o_0$   &  7.606$-$11  &  7.690$-$11  &  7.589$-$11  \\
   14  &    2s3p    &  $^3$P$^o_1$   &  8.132$-$09  &  8.135$-$09  &  8.339$-$09  \\
   15  &    2s3p    &  $^3$P$^o_2$   &  7.352$-$09  &  7.503$-$09  &  8.078$-$09  \\
   16  &    2s3p    &  $^1$P$^o_1$   &  8.084$-$09  &  8.087$-$09  &  8.276$-$09  \\
   17  &    2s3d    &  $^3$D$  _1$   &  3.338$-$11  &  3.329$-$11  &  3.295$-$11  \\
   18  &    2s3d    &  $^3$D$  _2$   &  3.339$-$11  &  3.330$-$11  &  3.296$-$11  \\
   19  &    2s3d    &  $^3$D$  _3$   &  3.341$-$11  &  3.332$-$11  &  3.298$-$11  \\
   20  &    2s3d    &  $^1$D$  _2$   &  5.270$-$11  &  5.322$-$11  &  5.400$-$11  \\
\hline  
\end{tabular}

\begin{flushleft}
{\small
GRASP1: Energies from the {\sc grasp} code for 166 level calculations \\
GRASP2: Energies from the {\sc grasp} code for 238 level calculations \\
}
\end{flushleft}

\end{table*}

\setcounter{table}{6}     
\begin{table*}      
\caption{Collision strengths for resonance transitions of  N~IV. $a{\pm}b \equiv$ $a\times$10$^{{\pm}b}$. See Table 1 for level indices. Complete table is available online as Supporting Information.}          
\begin{tabular}{rrlllllllr}                                                                                   
\hline                                                                                                                                                                                                             
\multicolumn{2}{c}{Transition} & \multicolumn{8}{c}{Energy (Ryd)}\\                                           
\hline                                                                                                        
  $i$ & $j$ &   10  &  15  &   20 & 25 & 30 & 35 & 40 & 45  \\                                                
\hline 
    1 &    2 &  1.178$-$02 &  6.958$-$03 &  4.491$-$03 &  3.131$-$03 &  2.317$-$03 &  1.806$-$03 &  1.449$-$03 &  1.318$-$03 \\
    1 &    3 &  3.535$-$02 &  2.089$-$02 &  1.349$-$02 &  9.412$-$03 &  6.968$-$03 &  5.438$-$03 &  4.369$-$03 &  3.975$-$03 \\
    1 &    4 &  5.886$-$02 &  3.477$-$02 &  2.244$-$02 &  1.564$-$02 &  1.157$-$02 &  9.023$-$03 &  7.240$-$03 &  6.582$-$03 \\
    1 &    5 &  6.292$+$00 &  7.242$+$00 &  7.928$+$00 &  8.462$+$00 &  8.902$+$00 &  9.282$+$00 &  9.617$+$00 &  9.976$+$00 \\
    1 &    6 &  2.418$-$04 &  1.257$-$04 &  7.227$-$05 &  4.522$-$05 &  3.020$-$05 &  2.110$-$05 &  1.583$-$05 &  1.238$-$05 \\
    1 &    7 &  7.228$-$04 &  3.751$-$04 &  2.149$-$04 &  1.340$-$04 &  8.904$-$05 &  6.185$-$05 &  4.611$-$05 &  3.587$-$05 \\
    1 &    8 &  1.207$-$03 &  6.295$-$04 &  3.631$-$04 &  2.284$-$04 &  1.537$-$04 &  1.085$-$04 &  8.242$-$05 &  6.545$-$05 \\
    1 &    9 &  1.423$-$01 &  1.372$-$01 &  1.344$-$01 &  1.328$-$01 &  1.319$-$01 &  1.313$-$01 &  1.311$-$01 &  1.320$-$01 \\
    1 &   10 &  3.503$-$02 &  3.359$-$02 &  3.174$-$02 &  2.997$-$02 &  2.833$-$02 &  2.658$-$02 &  2.463$-$02 &  2.233$-$02 \\
    ... & \\
    ... & \\
    ... & \\
\hline                                                                                                        
\end{tabular}                                                                                                 
\end{table*}                                                   

\section{Lifetimes}

In contrast to energy levels, there are no direct measurements of radiative rates to compare with theoretical results. However, the A-values are related to the lifetime $\tau$  as follows:

\begin{equation}  {\tau}_j = \frac{1}{{\sum_{i}^{}} A_{ji}} 
\end{equation} 
and several measurements for levels of N~IV are available in the literature. Furthermore, if  a single transition dominates the contributions then one can effectively obtain an `indirect' measurement of the A-value.  Therefore, in Table~1 we have also listed our calculated lifetimes, from both the GRASP1 and GRASP2 models. As noted earlier, A-values for E1 transitions are often larger in magnitude and hence dominate  the determination of $\tau$. However,  we have also  included the contributions from E2, M1 and M2 transitions, which can be  important for those levels which do not have any dominating  E1 connection.

Measurements of $\tau$ for levels of N~IV (up to 1990)  have been compiled by \cite{nl}, and are compared in Table~5 with our results from both the GRASP1 and GRASP2 models.  Both models yield almost the identical results for all the levels listed in this table, except one, i.e. 2s5p $^1$P$^o_1$. For this,  our GRASP2 value of $\tau$ is closer to the measurement.  There are several measurements for some levels and therefore we have listed the  range of values, with specific results  given in  table IIIa of \cite{nl}.  \cite{eng} have also measured $\tau$ for the 2s2p $^1$P$^o_1$ level which was not included by \cite{nl}. Their measured value of 0.425$\pm$0.015 ns is closer to the lower end of the range (0.44 -- 0.53~ns) listed by \cite{nl}, and is only larger than our calculation by 14\% (0.37~ns).  Similarly, for most of the levels listed in Table~5, there is reasonable agreement (within a few percent) between theory and measurements. However, there are two exceptions, namely 2p3p $^3$P and  2s4s $^3$S. For the former, the measured value of 7.83$\pm$0.08 ns by  \cite{des} is much larger than the  0.18$\pm$0.02 ns of \cite{buck} and our result of 0.183 ns, and hence appears to be incorrect. In the case of  2s4s $^3$S, our calculation of $\sim$3 ns is larger than the measurement (0.12$\pm$ 0.01 ns) of  \cite{buck} by a factor of 25.  However, the theoretical results are consistent over a period of time. For example, the dominating contributing E1 transitions are: 2s3p $^3$P$^o_{0,1,2}$--2s4s $^3$S$_1$ (i.e. 14/15/16--37) for which our A-values (from both GRASP1 and GRASP2) are 2.8$\times$10$^7$, 8.8$\times$10$^7$ and 1.6$\times$10$^8$ s$^{-1}$, respectively, whereas those calculated by \cite*{jat} and stored in the NIST database are 3.01$\times$10$^7$, 9.02$\times$10$^7$ and 1.50$\times$10$^8$  s$^{-1}$, respectively, i.e. agreeing to better than 10\% with our results. Similarly, the A-values of \cite{icft} for the corresponding transitions are  2.9$\times$10$^7$, 9.4$\times$10$^7$ and 1.76$\times$10$^8$ s$^{-1}$, respectively, again agreeing within 10\% with our calculations. Finally,  the f-value calculated for the 2s3p $^3$P$^o$--2s4s $^3$S transition by \cite{nh} is 0.014 whereas our result is 0.016. Therefore, we are confident of our (and other theoretical) results and suspect that the  $\tau$ measurements for the 2s4s $^3$S level, are in error.

 \cite{tff} have calculated A-values for transitions among the lowest 20 levels of Be-like ions, including N~IV. They did not report the corresponding $\tau$ values, but these are available on their website: {\tt http://nlte.nist.gov/MCHF/view.html}.   Our results are compared with their calculations  in  Table~6 and there are no discrepancies.

 \section{Collision strengths}

The collision strength for electron impact excitation  ($\Omega$), a symmetric and dimensionless quantity,  is related to the better-known parameter collision cross section ($\sigma_{ij}$), by a simple equation  (7) given in  \cite{tixix}. As stated  in section 1 (and our work on many Be-like ions),   we have adopted the  relativistic {\sc darc} code  (standard and parallelised versions)  for the scattering calculations. This code is based on the $jj$ coupling scheme and uses the  Dirac-Coulomb Hamiltonian in an R-matrix approach. Two sets of calculations have been performed, one (DARC1) with 166 levels of the GRASP1 model, and another (DARC2) with 238 levels of GRASP2.  The DARC1 calculations for N~IV are larger than those performed for other Be-like ions with 13 $\le$ Z $\le$ 32 (see \cite{belike} and references therein) but are similar to those for C~III \citep{c3}. Our DARC2 calculations are even larger, and exactly match in size with those of \cite{icft}. For N~IV, the  adopted R-matrix radius (Ra) and the number of  continuum orbitals for each channel angular momentum (NRANG2) are 21.6 au and 55, for DARC1. Correspondingly,  the maximum number of channels generated for a partial wave is 828 which leads to  the Hamiltonian (H) matrix size of 45~714. For the DARC2 calculations, Ra and NRANG2 are 35.2 au and 88, respectively. The maximum number of channels generated in this calculations is 1116 and the corresponding H-size is 98~478. To achieve  convergence of  $\Omega$ for most transitions and at (almost) all energies,  all partial waves with angular momentum $J \le$ 40.5  have been included in both calculations. Furthermore, in both, the contribution of  higher neglected partial waves has been included through  a top-up procedure, based on the Coulomb-Bethe  \citep*{ab} and  geometric series  approximations for allowed and forbidden transitions, respectively. Thus care has been taken to ensure the accuracy of our calculated values of $\Omega$, as for other Be-like ions. Finally, values of  $\Omega$ have been calculated up to energies of  35  and 45~Ryd for DARC1 and DARC2, respectively. Subsequently, the values of effective collision strength $\Upsilon$ (see section 6) are calculated up to T$_e$ = 1.5$\times$10$^{6}$ K in DARC1, and up to T$_e$ = 2.0$\times$10$^{6}$ K in DARC2. The temperature  of  maximum abundance in ionisation equilibrium for N~IV is only 1.26 $\times$10$^{5}$ K  \citep*{pb}, and hence  both calculations should cover all plasma applications.

\begin{figure*}
\includegraphics[angle=90,width=0.9\textwidth]{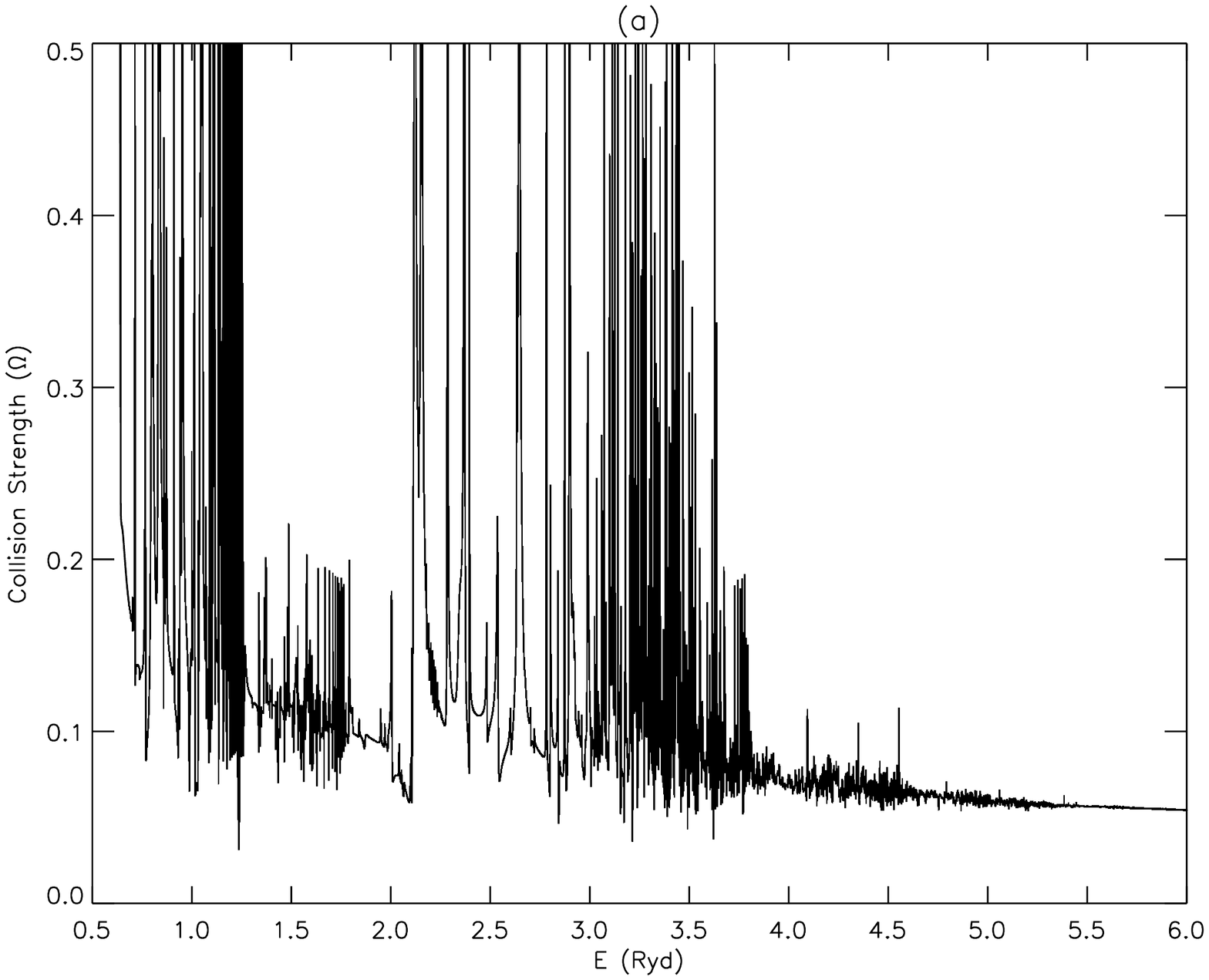}
 \caption{Collision strengths for the (a) 1--3 (2s$^2$ $^1$S$_0$--2s2p $^3$P$^o_1$), (b) 1--10 (2s$^2$ $^1$S$_0$--2p$^2$ $^1$S$_0$) and (c) 3--5 (2s2p $^3$P$^o_1$--2s2p $^1$P$_1^o$) transitions of N~IV.}
 \end{figure*}
 
\setcounter{figure}{1}
 \begin{figure*}
\includegraphics[angle=90,width=0.9\textwidth]{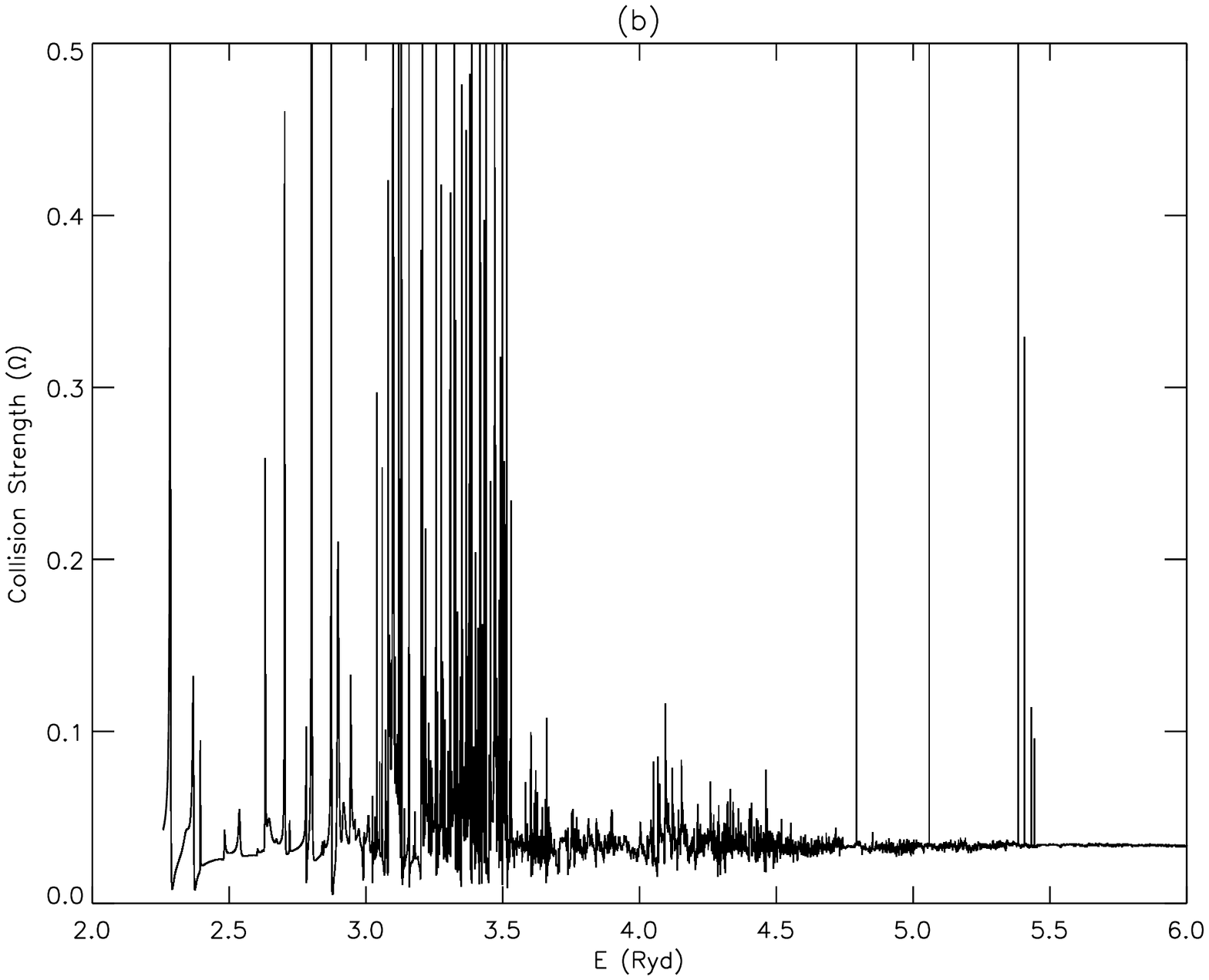}
\vspace{-1.0cm}
 \caption{ ..... continued.}
 \end{figure*}
 
  \vspace*{0.5 cm} 
 
 \setcounter{figure}{1} 
 \begin{figure*}
\includegraphics[angle=90,width=0.9\textwidth]{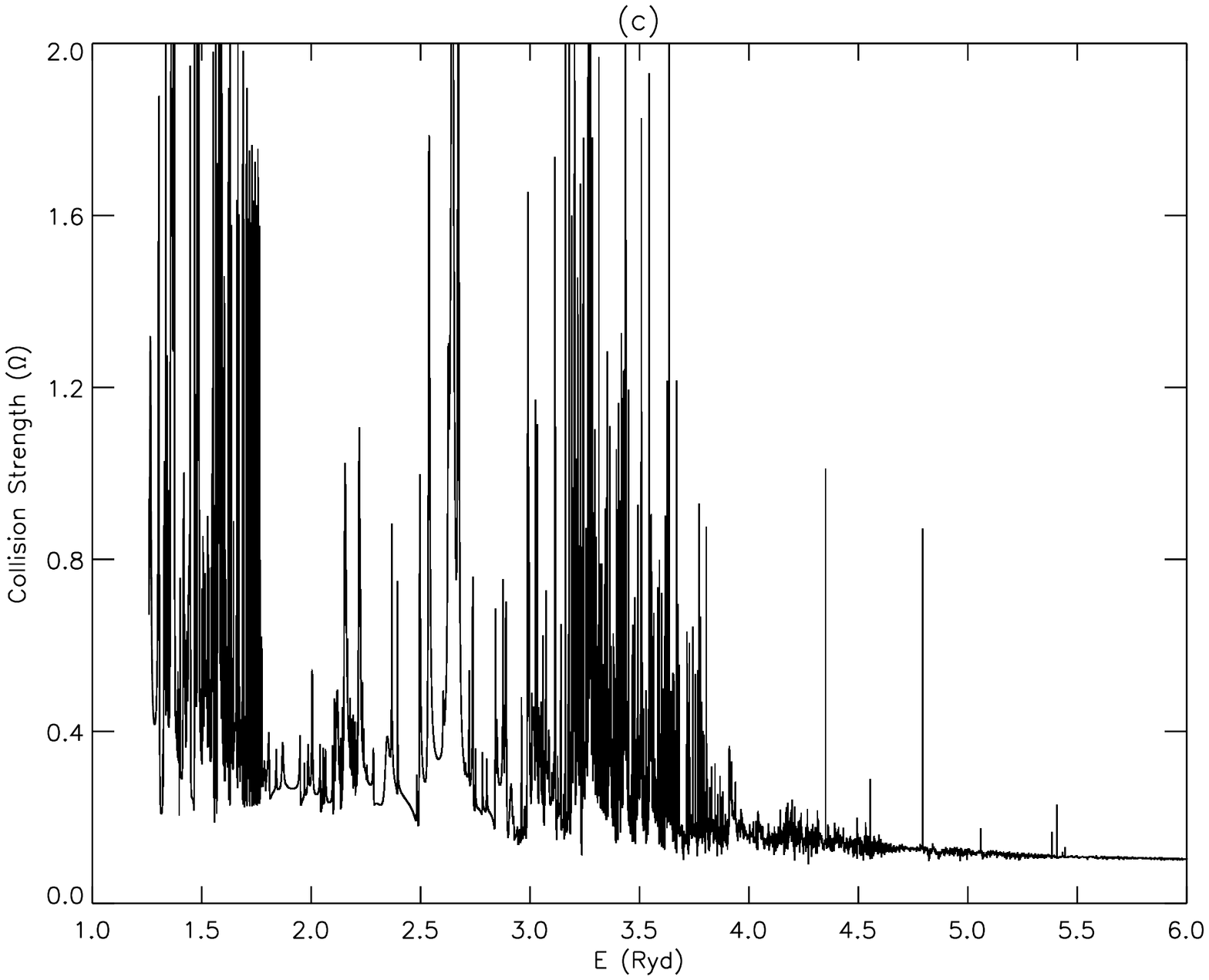}
 \vspace{-1.5cm}
 \caption{ ..... continued.}
 \end{figure*}

Unfortunately, prior theoretical (or experimental) data for $\Omega$ are not available for comparison  with  our results. At energies above thresholds, $\Omega$ varies smoothly and therefore  in Table~7 we list our values of $\Omega$ for resonance transitions of  N~IV at  energies in the 10--45 Ryd range. These $\Omega$ are from our DARC2 calculations and will hopefully be useful for future comparison with experimental and other theoretical results. However, a comparison of $\Omega$ made with the DARC1 calculations, for the lowest 78 levels, shows a satisfactory agreement within $\sim$10\%, except for the 1--58 (2s$^2$~$^1$S$_0$ -- 2p3p $^1$S$_0$) transition for which the differences are 50\%. By contrast, the  threshold energy region is often dominated by numerous closed-channel (Feshbach) resonances, as shown in fig.~2 of  \cite{icft} for four transitions, namely 1--3 (2s$^2$ $^1$S$_0$ -- 2s2p $^3$P$^o_{1}$), 1--4 (2s$^2$ $^1$S$_0$ -- 2s2p $^3$P$^o_{2}$), 1--5 (2s$^2$ $^1$S$_0$ -- 2s2p $^1$P$^o_{1}$) and 1--9 (2s$^2$ $^1$S$_0$ -- 2p$^2$ $^1$D$_{2}$) for three Be-like ions,  C~III, Mg~IX and Fe~XXIII.  Similarly, resonances have been shown by us \citep{tixix} for six transitions of Ti~XIX and two of C~III \citep{c3}. Regarding N~IV, \cite{rams} have shown resonances for four transitions, namely 2s$^2$ $^1$S--2s2p $^3$P$^o$, 2s$^2$ $^1$S--2p$^2$ $^1$S,    2s$^2$ $^1$S--2s3p $^1$P$^o$ and 2s2p $^3$P$^o$--2s2p $^1$P$^o$. We discuss these resonances in the next section.  

\section{Effective collision strengths}

Since $\Omega$ does not vary smoothly with energy in the thresholds region,  its values are averaged  over a suitable distribution of electron velocities  to determine the  'effective' collision strength ($\Upsilon$). For most plasma modelling applications,  a Maxwellian  distribution of electron velocities is assumed, to obtain $\Upsilon$ from:

\begin{equation}
\Upsilon(T_e) = \int_{0}^{\infty} {\Omega}(E) \, {\rm exp}(-E_j/kT_e) \,d(E_j/{kT_e}),
\end{equation}
where $k$ is Boltzmann constant, T$_e$  the electron temperature in K, and E$_j$  the electron energy with respect to the final (excited) state. Once the value of $\Upsilon$ is
known the corresponding results for the excitation q(i,j) and de-excitation q(j,i) rates can be easily obtained from the following equations:

\begin{equation}
q(i,j) = \frac{8.63 \times 10^{-6}}{{\omega_i}{T_e^{1/2}}} \Upsilon \, {\rm exp}(-E_{ij}/{kT_e}) \hspace*{1.0 cm}{\rm cm^3s^{-1}}
\end{equation}
and
\begin{equation}
q(j,i) = \frac{8.63 \times 10^{-6}}{{\omega_j}{T_e^{1/2}}} \Upsilon \hspace*{1.0 cm}{\rm cm^3 s^{-1}},
\end{equation}
where $\omega_i$ and $\omega_j$ are the statistical weights of the initial ($i$) and final ($j$) states, respectively, and E$_{ij}$ is the transition energy.

Often, the contribution of resonances over the background  collision strengths ($\Omega_{\rm B}$)  is significant (by up to an order of magnitude or even more), but this strongly depends on the type of transition, such as forbidden, semi-forbidden and inter-combination. Similarly, values of $\Upsilon$ are affected by the resonances more at low(er)  temperatures than at higher ones. Therefore,  it is important to resolve resonances  in a fine energy mesh  so that their contribution can be properly taken into account. If the energy mesh is too broad then either some of the resonances may be missed (and subsequently $\Upsilon$ may be underestimated) or one may get two consecutive peaks, leading to an overestimation of $\Upsilon$. On the other hand, if the energy mesh is too fine and the resonances are not too dense, as in the case of C~III \citep{c3}, then one may unnecessarily spend  time in computational effort  without gaining any advantage. Therefore, a careful balance is required in determining the mesh size, and this is important considering the large size of the H matrix.

\setcounter{table}{7}                                                                                                                       
\begin{table*}                                                                                                                              
\caption{Effective collision strengths for transitions in  N IV. $a{\pm}b \equiv a{\times}10^{{\pm}b}$. Complete table is available online as Supporting Information.}                                    
\begin{tabular}{rrlllllllllll}                                                                                                              
\hline                                                                                                                                      
\multicolumn {2}{c}{Transition} & \multicolumn{10}{c}{Temperature (log T$_e$, K)}\\                                                         
\hline                                                                                                                                      
$i$ & $j$ &     4.50  &    4.70  &    4.90   &   5.10  &    5.30  &    5.50  &   5.70  &    5.90 &  6.10  &  6.30  \\                       
\hline                                                                                                                                                       
    1 &    2 &  1.010$-$01 &  8.868$-$02 &  7.776$-$02 &  6.804$-$02 &  5.905$-$02 &  5.022$-$02 &  4.153$-$02 &  3.331$-$02 &  2.588$-$02 &  1.945$-$02 \\
    1 &    3 &  2.969$-$01 &  2.620$-$01 &  2.307$-$01 &  2.025$-$01 &  1.761$-$01 &  1.500$-$01 &  1.242$-$01 &  9.967$-$02 &  7.748$-$02 &  5.826$-$02 \\
    1 &    4 &  4.745$-$01 &  4.236$-$01 &  3.761$-$01 &  3.321$-$01 &  2.901$-$01 &  2.479$-$01 &  2.056$-$01 &  1.653$-$01 &  1.286$-$01 &  9.674$-$02 \\
    1 &    5 &  3.066$+$00 &  3.146$+$00 &  3.258$+$00 &  3.417$+$00 &  3.642$+$00 &  3.954$+$00 &  4.373$+$00 &  4.909$+$00 &  5.523$+$00 &  6.001$+$00 \\
    1 &    6 &  2.157$-$03 &  2.081$-$03 &  2.089$-$03 &  2.105$-$03 &  2.011$-$03 &  1.777$-$03 &  1.459$-$03 &  1.131$-$03 &  8.381$-$04 &  5.996$-$04 \\
    1 &    7 &  6.481$-$03 &  6.243$-$03 &  6.258$-$03 &  6.291$-$03 &  6.002$-$03 &  5.297$-$03 &  4.347$-$03 &  3.368$-$03 &  2.497$-$03 &  1.786$-$03 \\
    1 &    8 &  1.088$-$02 &  1.045$-$02 &  1.045$-$02 &  1.048$-$02 &  9.989$-$03 &  8.816$-$03 &  7.237$-$03 &  5.610$-$03 &  4.160$-$03 &  2.978$-$03 \\
    1 &    9 &  1.842$-$01 &  1.887$-$01 &  1.931$-$01 &  1.949$-$01 &  1.918$-$01 &  1.846$-$01 &  1.753$-$01 &  1.659$-$01 &  1.568$-$01 &  1.460$-$01 \\
    1 &   10 &  4.429$-$02 &  4.453$-$02 &  4.585$-$02 &  4.614$-$02 &  4.487$-$02 &  4.272$-$02 &  4.049$-$02 &  3.846$-$02 &  3.645$-$02 &  3.374$-$02 \\
    ... & \\
    ... & \\
    ... & \\
\hline                                                                                                                                      
\end{tabular}                                                                                                                               
\end{table*}

 \begin{figure*}
\includegraphics[angle=-90,width=0.9\textwidth]{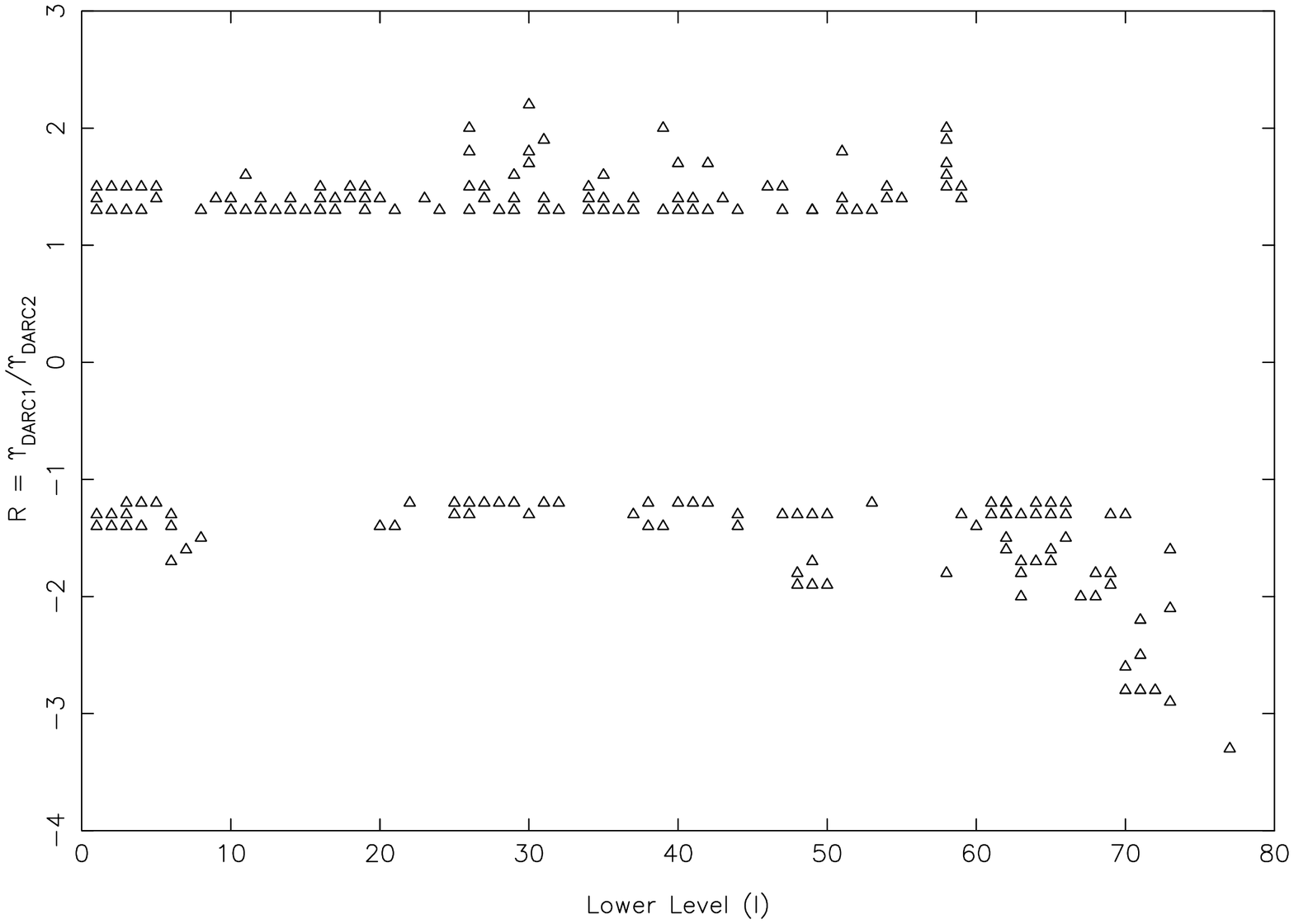}
 \vspace{-1.5cm}
 \caption{Comparison of DARC1 and DARC2 $\Upsilon$ for transitions of N~IV at T$_e$ = 1.6$\times$10$^{5}$ K. Negative R values indicate that $\Upsilon_{\rm DARC2}$ $>$ $\Upsilon_{\rm DARC1}$ and only those transitions are shown which differ by over 20\%.}
 \end{figure*}

 \begin{figure*}
\includegraphics[angle=-90,width=0.9\textwidth]{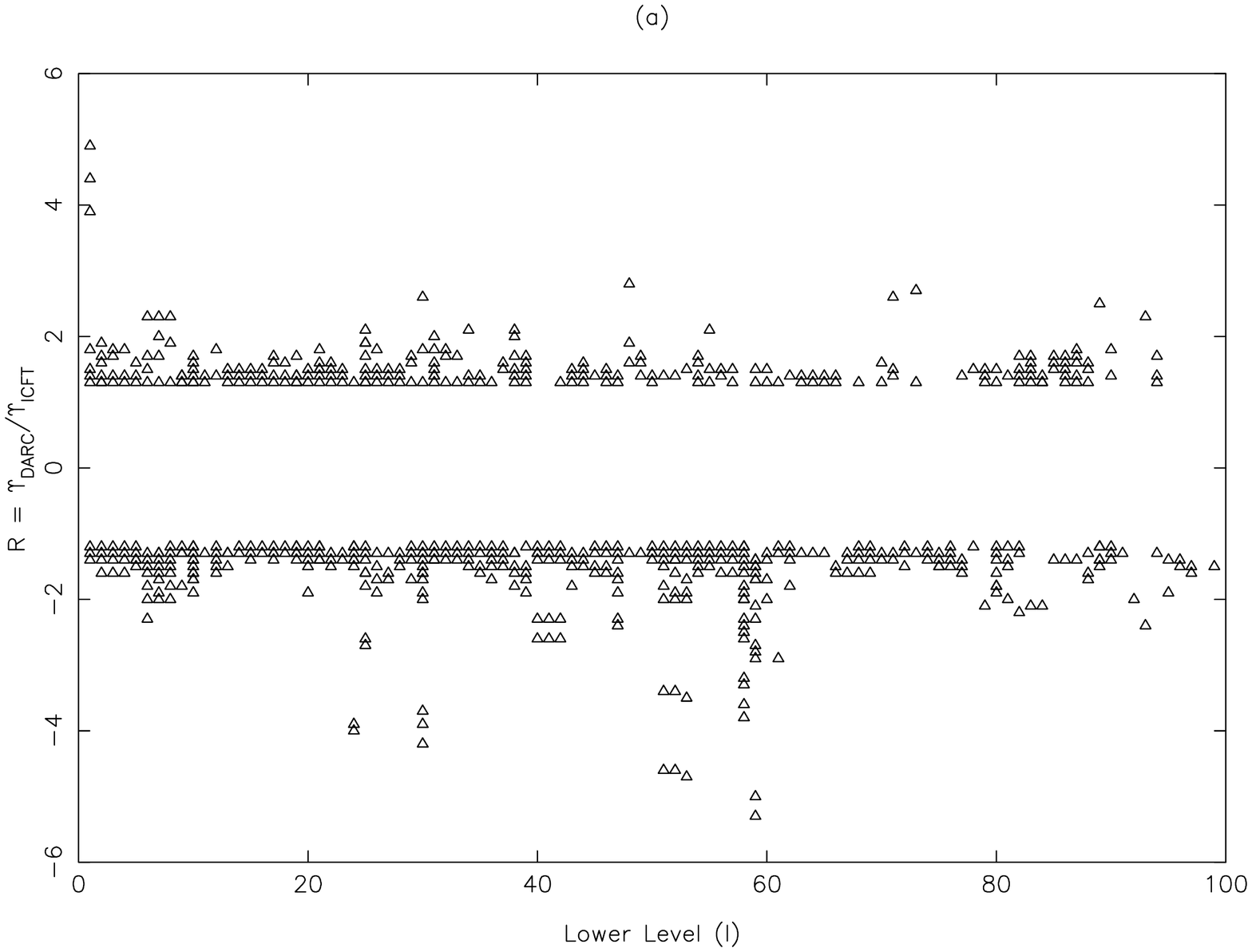}
 \vspace{-1.5cm}
 \caption{Comparison of DARC2 and ICFT values of $\Upsilon$ for transitions of N~IV at (a) T$_e$ = 3.2$\times$10$^{3}$ K, (b) T$_e$ = 1.6$\times$10$^{5}$ K,  and (c and d) T$_e$ = 1.6$\times$10$^{6}$ K. Negative R values plot  $\Upsilon_{\rm ICFT}$/$\Upsilon_{\rm DARC2}$ and indicate that $\Upsilon_{\rm ICFT}$ $>$ $\Upsilon_{\rm DARC2}$. Only those transitions are shown which differ by over 20\%.}
 \end{figure*}

\setcounter{figure}{3} 
 \begin{figure*}
\includegraphics[angle=-90,width=0.9\textwidth]{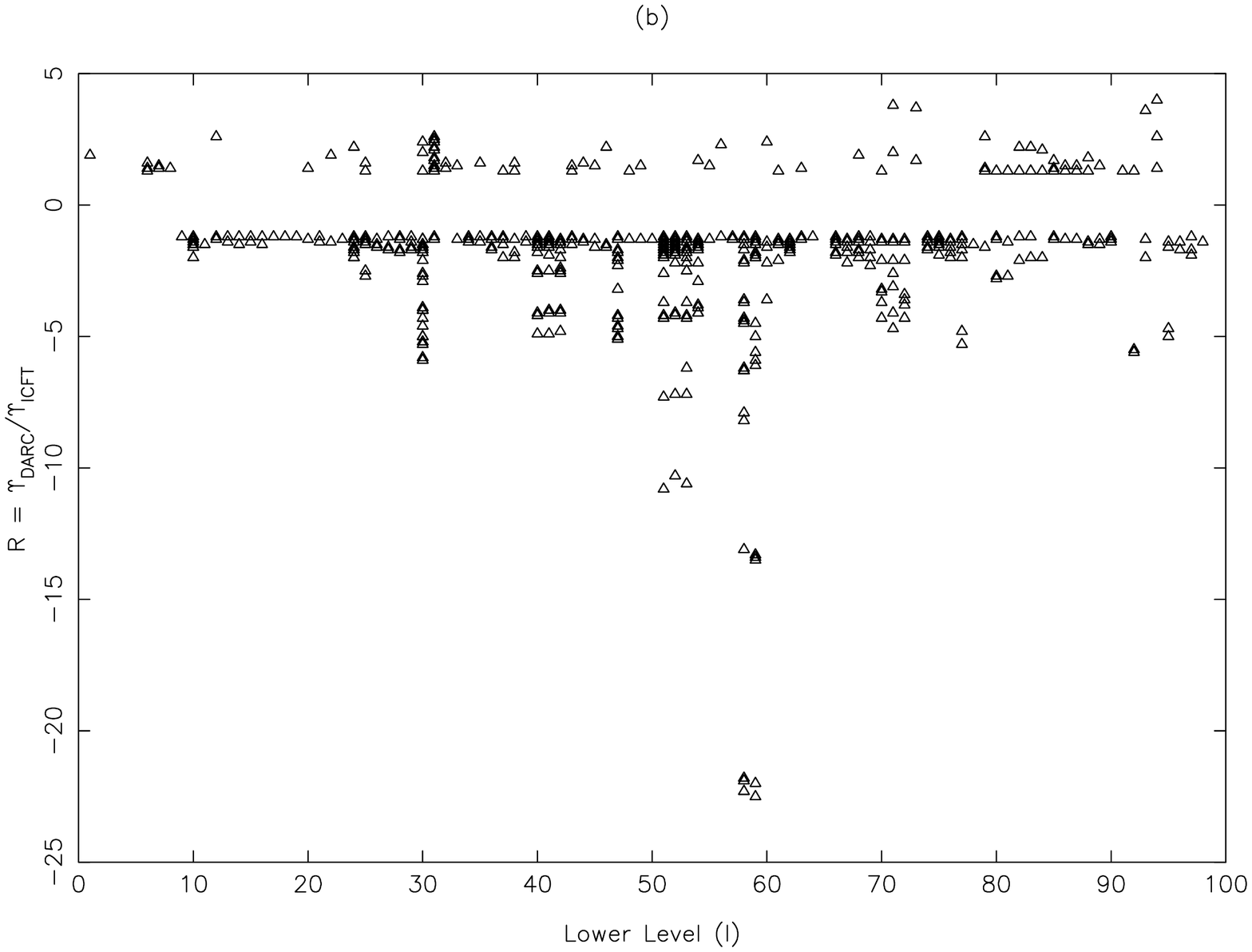}
 \vspace{-1.5cm}
 \caption{ ..... continued.}
 \end{figure*}

\setcounter{figure}{3} 
 \begin{figure*}
\includegraphics[angle=-90,width=0.9\textwidth]{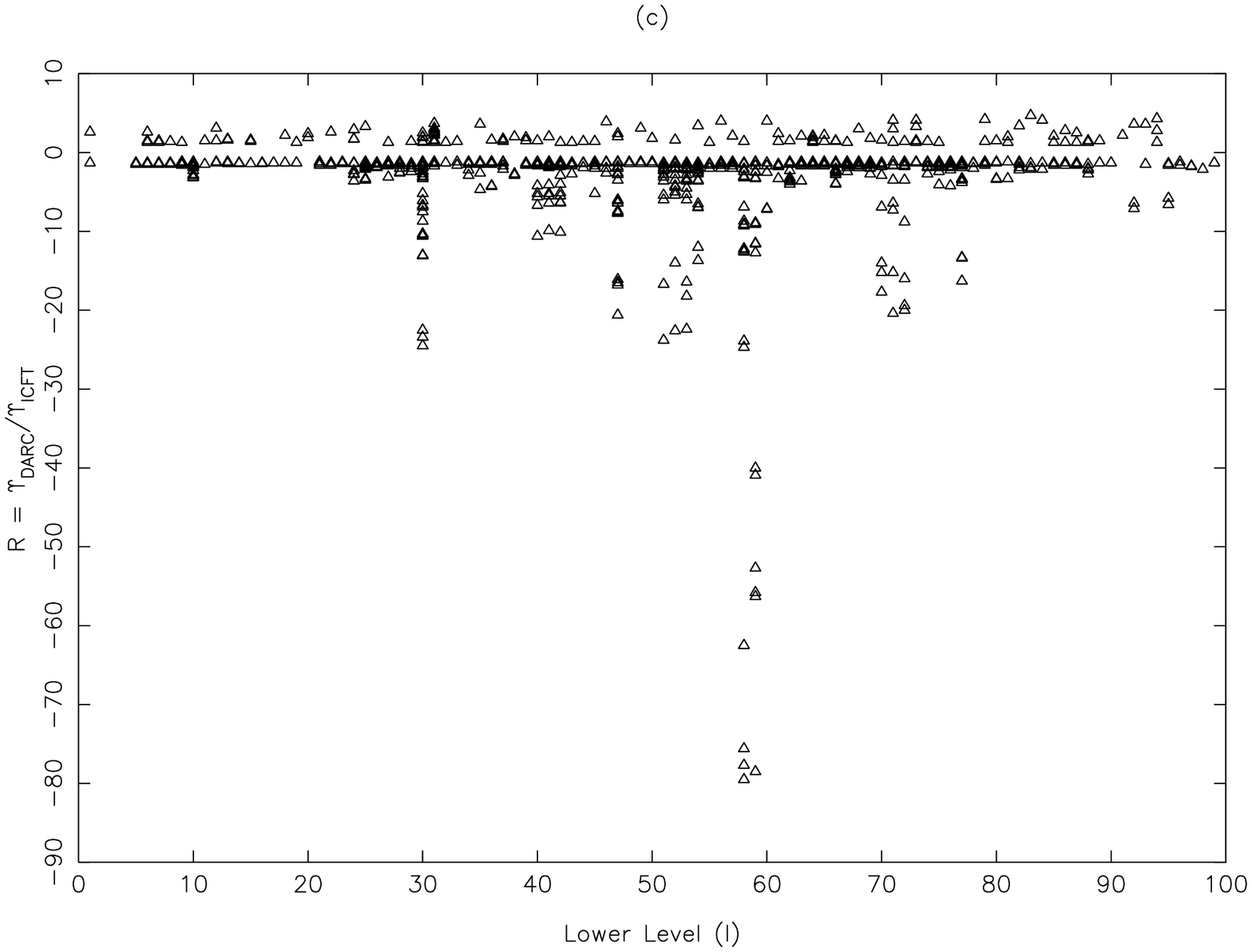}
 \vspace{-1.5cm}
 \caption{ ..... continued.}
 \end{figure*}

\setcounter{figure}{3} 
 \begin{figure*}
\includegraphics[angle=-90,width=0.9\textwidth]{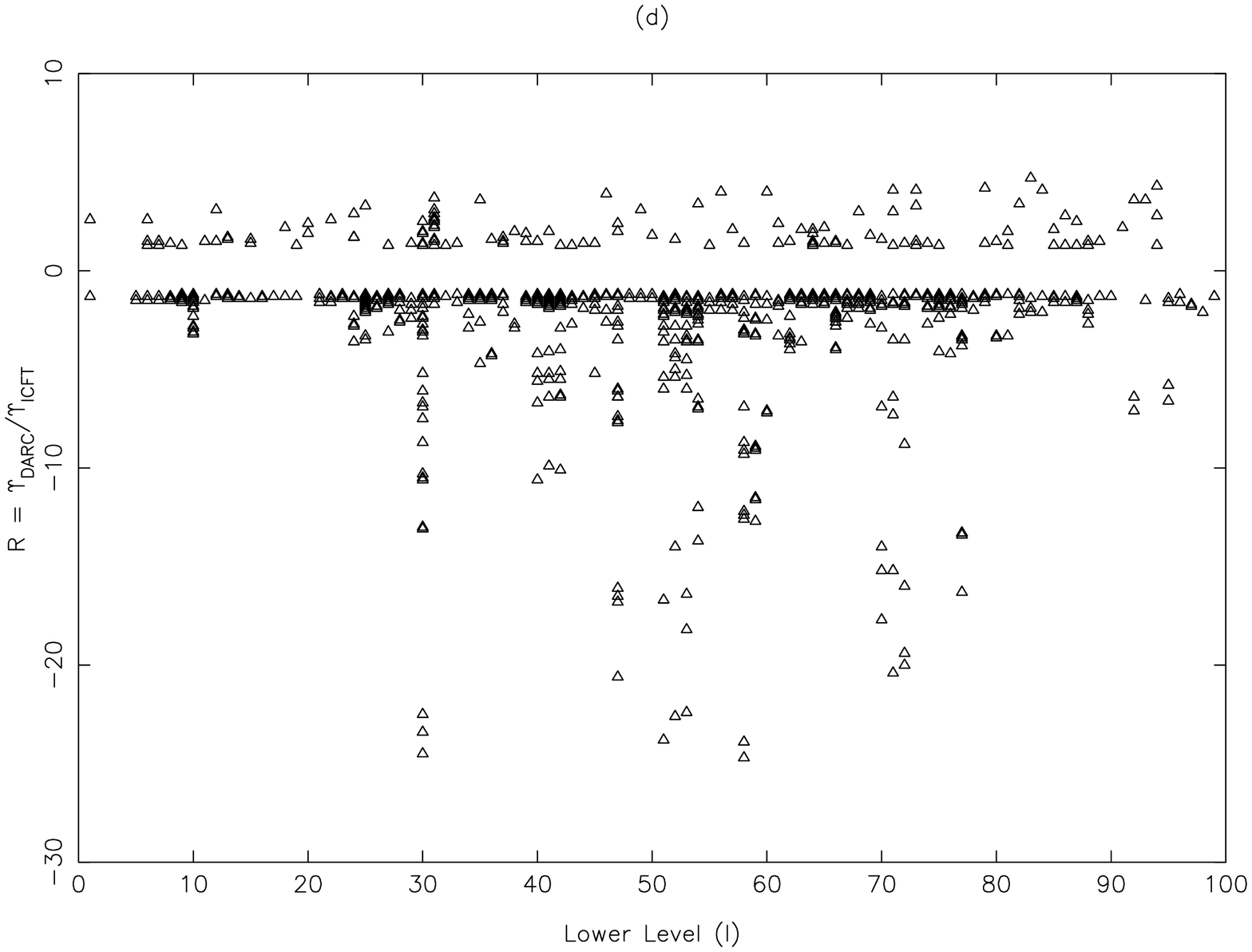}
 \vspace{-1.5cm}
 \caption{ ..... continued.}
 \end{figure*}

\setcounter{figure}{4} 
 \begin{figure*}
\includegraphics[angle=-90,width=0.9\textwidth]{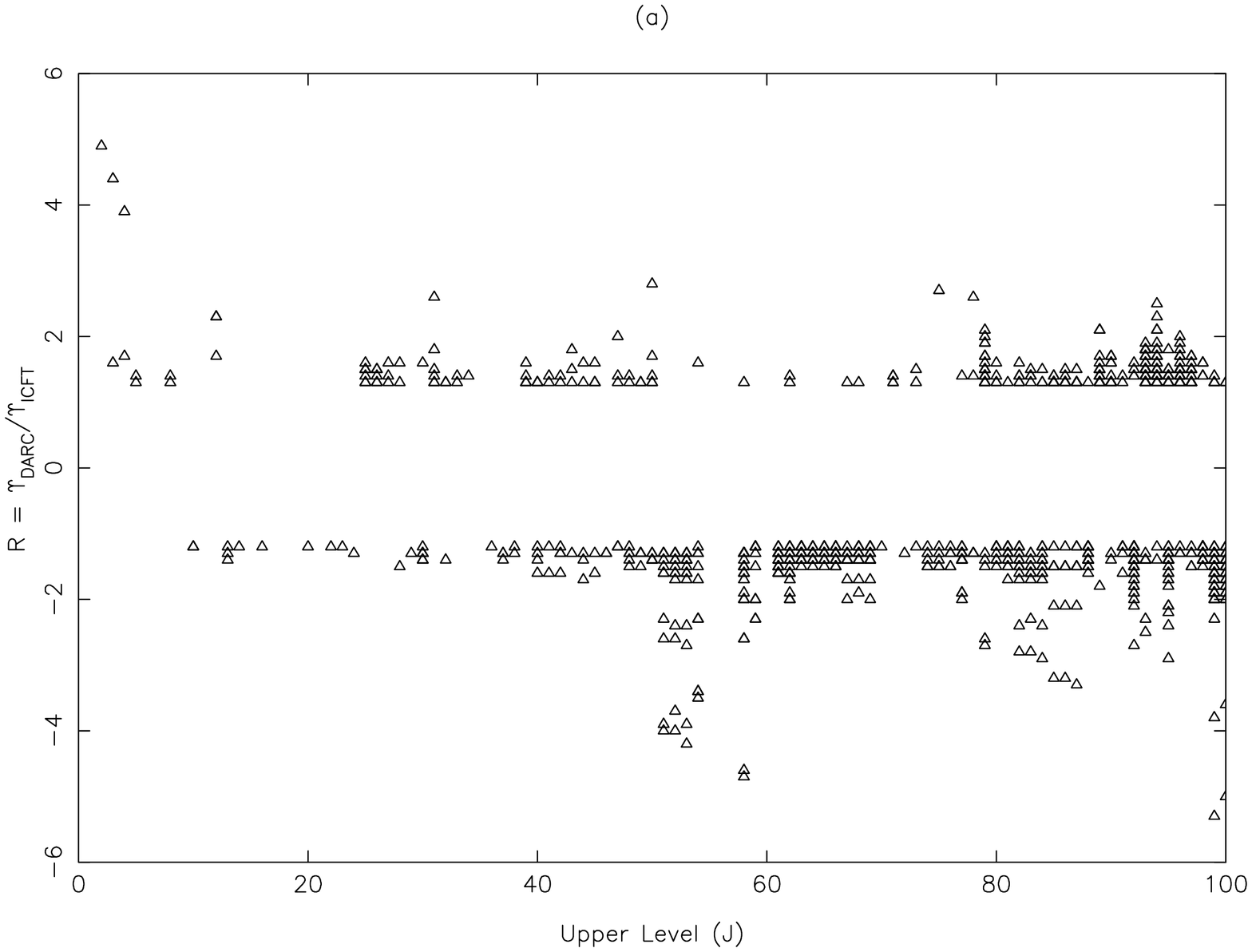}
 \vspace{-1.5cm}
 \caption{Comparison of DARC2 and ICFT $\Upsilon$ for transitions of N~IV at (a) T$_e$ = 3.2$\times$10$^{3}$ K, (b) T$_e$ = 1.6$\times$10$^{5}$ K,  and (c and d) T$_e$ = 1.6$\times$10$^{6}$ K. Negative R values plot  $\Upsilon_{\rm ICFT}$/$\Upsilon_{\rm DARC2}$ and indicate that $\Upsilon_{\rm ICFT}$ $>$ $\Upsilon_{\rm DARC2}$. Only those transitions are shown which differ by over 20\%.}
 \end{figure*}

\setcounter{figure}{4} 
 \begin{figure*}
\includegraphics[angle=-90,width=0.9\textwidth]{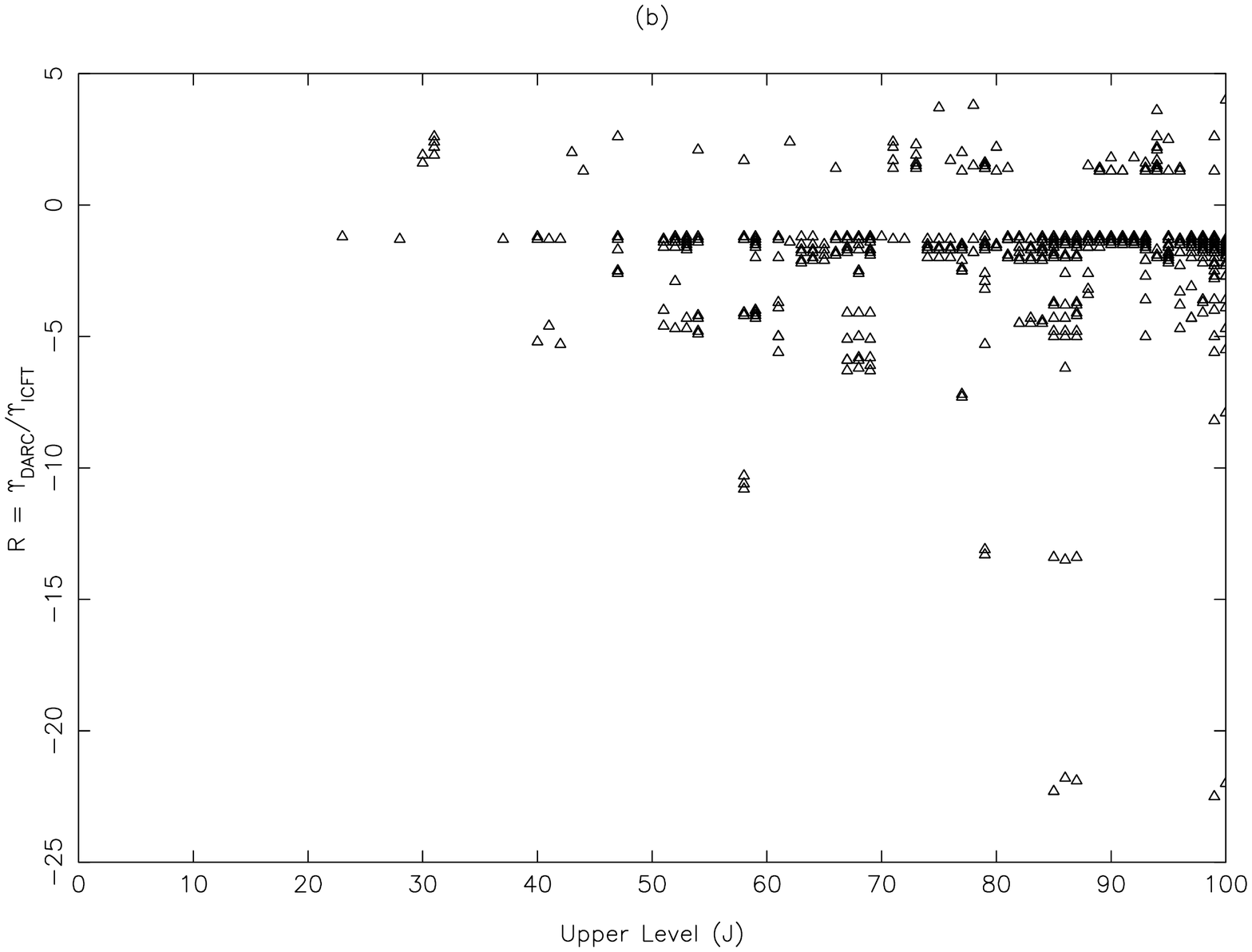}
 \vspace{-1.5cm}
 \caption{ ..... continued.}
 \end{figure*}

\setcounter{figure}{4} 
 \begin{figure*}
\includegraphics[angle=-90,width=0.9\textwidth]{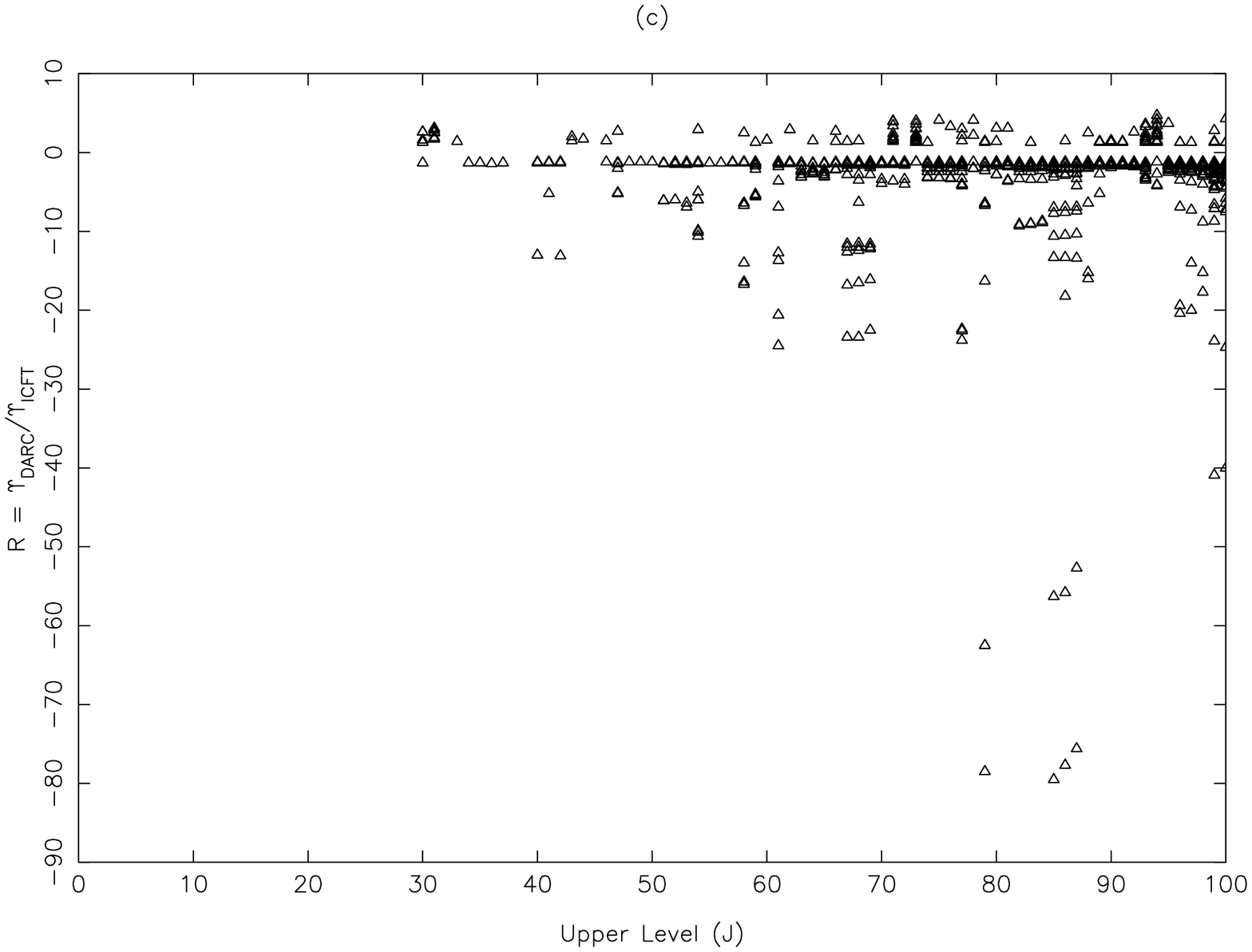}
 \vspace{-1.5cm}
 \caption{ ..... continued.}
 \end{figure*}

\setcounter{figure}{4} 
 \begin{figure*}
\includegraphics[angle=-90,width=0.9\textwidth]{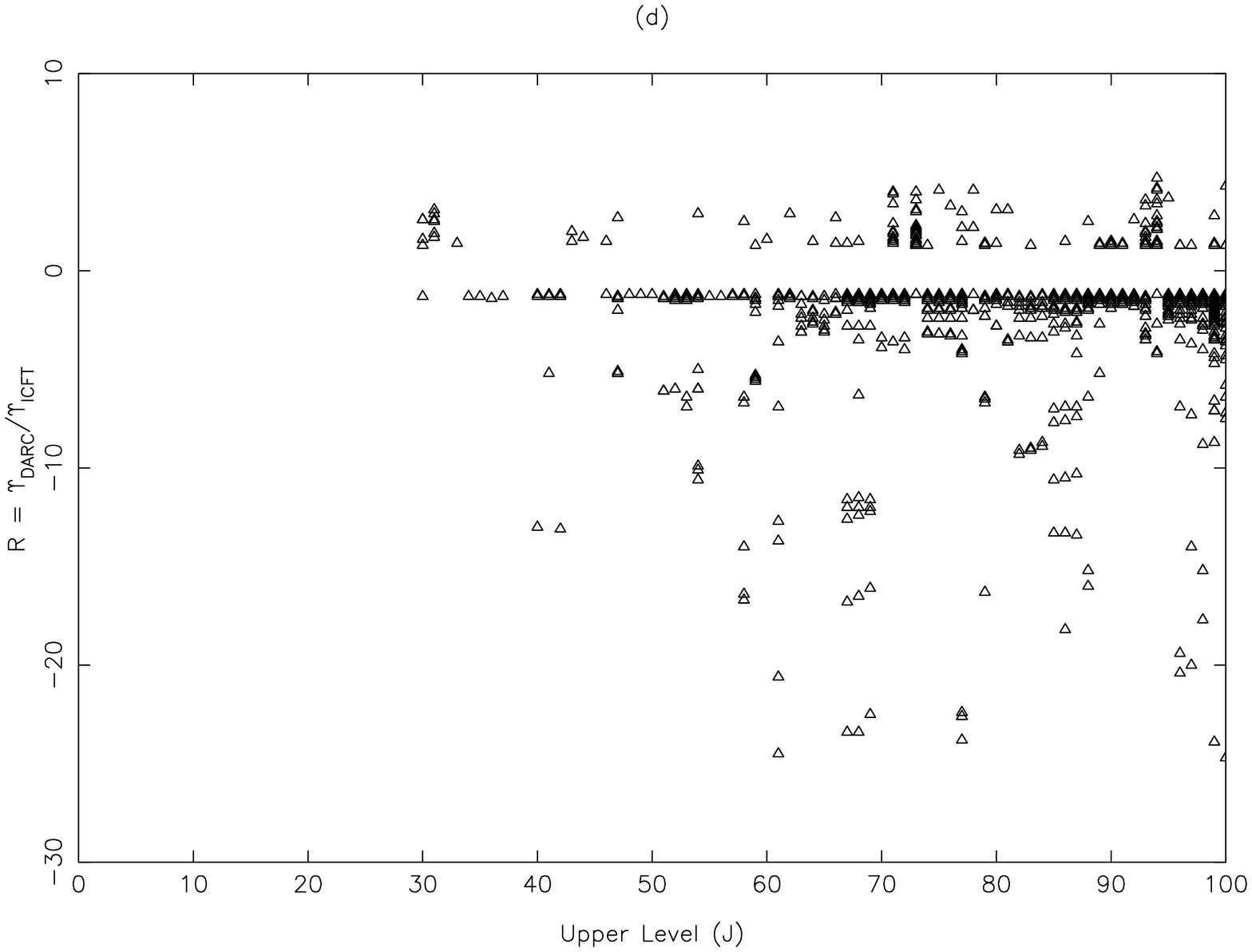}
 \vspace{-1.5cm}
 \caption{ ..... continued.}
 \end{figure*}

Since we want to resolve discrepancies between our calculations of $\Upsilon$ and those by \cite{icft}, we have performed two full calculations, i.e. DARC1 and DARC2. Our DARC1 calculations are similar to those for C~III \citep{c3}, i.e. the energy resolution ($\Delta$E) is generally 0.001~Ryd, although in a few small energy ranges it is 0.002~Ryd. Resonances have been resolved at a total of 3622 energies in the thresholds region. By comparison, our DARC2 calculations are much more extensive because over most of the energy range $\Delta$E is as small as 0.000~045~Ryd. Thus the DARC2 calculations have been performed at over 52~000 energies. Clearly, our DARC1 calculations are (comparatively much) coarser, but have still taken several months to compute. Similarly, the DARC2 calculations have also taken several months in spite of using a parallelised version of the code for this work, as  this calculation is much larger, in both the size of the H matrix as well as the energy resolution.

Before we discuss our results of $\Upsilon$, in Fig. 2 (a, b and c) we show resonances (from the DARC2 calculations) for three transitions, namely 2s$^2$ $^1$S$_0$--2s2p $^3$P$^o_1$ (1--3), 2s$^2$ $^1$S$_0$--2p$^2$ $^1$S$_0$ (1--10) and 2s2p $^3$P$^o_1$--2s2p $^1$P$_1^o$ (3--5). The first  is an inter-combination transition whereas the other two are  forbidden. Similar dense  resonances have been detected for many more transitions. Our calculated values of $\Upsilon$ (DARC2) are listed in Table~8 over a wide temperature range up to 2$\times$10$^{6}$ K, suitable for application to a wide range of astrophysical (and laboratory) plasmas. Data at any intermediate temperature can be easily interpolated, because (unlike $\Omega$) $\Upsilon$ is a slowly varying function of T$_e$. Our corresponding results from DARC1 are not listed  here but can be obtained from the first author (KMA) on request. As  noted in section~1, the most recent and extensive  corresponding data  available for $\Upsilon$ are those by \cite{icft}.  Similar to our work, they have adopted the (semi-relativistic) R-matrix code,  resolved resonances in a fine energy mesh ($\sim$0.000~09~Ryd),  averaged $\Omega$ over a Maxwellian distribution of electron velocities, and  reported results for (fine-structure) transitions among 238 levels, over a wide range of electron temperature up to 3.2$\times$10$^7$ K. However, they divided their calculations into two parts. For $J \le$ 11.5 they performed electron {\em exchange} calculations but neglected this for higher $J$ values. This should not  affect the accuracy of the calculations. However, for higher $J$ their $\Delta$E was coarser (0.009~Ryd), which sometimes may be a limiting source of  error in calculating $\Upsilon$. By contrast, our calculations have the same energy resolution for all partial waves. Similarly, they calculated values of $\Omega$ up to an energy of about 17~Ryd, and beyond that dipole and Born limits were used to extrapolate results up to infinite energy, whereas we have determined $\Omega$ up to 35 and 45~Ryd for DARC1 and DARC2, sufficient to calculate $\Upsilon$ in the temperature range of interest. This is (perhaps) a crucial difference between the two methodologies and hence a major source of discrepancy, as already noted in \cite{belike}. To address the discrepancies in $\Upsilon$, we now undertake  detailed comparisons between different sets of results.

We first compare our values of $\Upsilon$ from DARC1 and DARC2, to test the conclusion of \cite{al10} that differences in atomic structure (i.e. the size of a calculation) can give rise to the large discrepancies  noted by them. We confine this comparison to transitions among the lowest 78 levels, because all calculations have the same ordering for these, as seen in Table~1. We have made these comparisons at three temperatures of: TE1=3.2$\times$10$^3$, TE2=1.6$\times$10$^5$ and TE3=8.0$\times$10$^5$~K. TE1 is the lowest temperature at which \cite{icft} have calculated their results and TE2 is closer to the most appropriate value for astrophysical applications \citep{pb}. At TE1, among the lowest 78 levels (3003 transitions), the $\Upsilon$ from DARC1 and DARC2 differ by over 20\% for 21\% of the transitions. For most,  $\Upsilon_{DARC2}$ are larger, generally within a factor of two, but for about 1\% of the transitions the discrepancies are up to a factor of 10. These differences are clearly understandable, mainly because (i) this is a very low temperature and hence sensitive to the position and magnitude of resonances, (ii) our $\Delta$E in DARC1 is coarse (0.001~Ryd equivalent to 158~K) and hence not suitable for calculations at such low temperatures, and (iii) our DARC2 calculations include more resonances.

A similar comparison at TE2  shows discrepancies for only 8\% of the transitions, mostly within a factor of two as shown in Fig. 3. For about half of the transitions, $\Upsilon_{DARC2}$ $>$ $\Upsilon_{DARC1}$, and the reverse is true for the rest. In fact, all those transitions which show (comparatively) larger discrepancies (less than a factor of 4) belong to levels higher than 70, and this is clearly due to the inclusion of resonances  from additional levels in DARC2. A similar conclusion applies at TE3, as for only 7\% of the transitions are there differences of  over 20\%, and almost all agree within a factor of two. Indeed, if the same fine(r) energy resolution had been adopted in DARC1, then the differences between the two sets of $\Upsilon$ might have been even less. Therefore, our conclusion is clearly different (see \cite{c3} and particularly their fig. 3) from those of \cite{al10}. A larger calculation certainly improves the accuracy of the calculated $\Upsilon$, but for most transitions (and particularly at temperatures of relevance) the discrepancies  are generally within $\sim$20\%. We now compare our values of $\Upsilon$ from DARC2 with the ICFT results of \cite{icft}. 

In Fig. 4 (a, b and c) we show the ratio R = $\Upsilon_{DARC2}$/$\Upsilon_{ICFT}$ (negative values  plot $\Upsilon_{ICFT}$/$\Upsilon_{DARC2}$ and indicate $\Upsilon_{ICFT}$ $> \Upsilon_{DARC2}$)  for all transitions among the lowest 100 levels of N~IV at three temperatures of TE1=3.2$\times$10$^3$, TE2=1.6$\times$10$^5$ and TE3=1.6$\times$10$^6$~K. The ratio R is shown as a function of transitions from {\em lower} levels I. At TE1, values of $\Upsilon$ for  about 22\% of the 4950 transitions differ by over 20\%, and for a majority of these the $\Upsilon_{ICFT}$ are larger (by up to a factor of 6). For transitions for which $\Upsilon_{DARC2}$ are larger the factor is generally below 3, except for three -- see Fig. 4a. Considering the fine energy resolution in both calculations and the inclusion of the same number of levels, such discrepancies are not expected. At TE2, a more relevant temperature for plasma modelling applications, the discrepancies are even worse because the $\Upsilon_{ICFT}$ are larger by up to a factor of 25 in most cases -- see Fig. 4b. This comparison, although similar to the one shown and discussed earlier \citep{belike, c3} for other Be-like ions, we believe, calls into question  the reliability of the calculations by \cite{icft}. TE3 corresponds to $\sim$10~Ryd, and therefore the contribution of resonances should not be as dominant as at lower temperatures (note that the highest threshold considered is at 6.1~Ryd -- see Table~1). Therefore, one would expect a (comparatively) better agreement between  $\Upsilon_{DARC2}$ and $\Upsilon_{ICFT}$. Unfortunately, the discrepancies become even greater than at lower T$_e$, as shown in Fig. 4c, because $\Upsilon_{ICFT}$ are larger by up to a factor of 80 in some instances. To obtain a clearer view of the discrepancies we show these again in Fig. 4d, in which the negative vertical scale has been reduced to 30. The $\Upsilon$ of \cite{icft} appear to be anomalous for many transitions and over a wide range of temperatures. We discuss these further below.

\cite{al10} have argued that instead of the lower levels I in Fig. 4, one should use the {\em upper} levels J to obtain a better comparison of the two calculations, because in a larger calculation the transitions among the lower levels should not be (much) affected -- see their fig. 5b. That should indeed be the case, although it does not apply in the present instance because both calculations are of the same size. Nevertheless, in Fig. 5 (a, b, c and d) we show similar comparisons as those in Fig. 4, but replacing I with J. Only transitions among the lowest $\sim$50 levels show a reasonably satisfactory agreement, and discrepancies are very large for those belonging to higher levels. This is extremely unsatisfactory, considering that there are 188 levels above the lowest 50. Since both calculations have the same size of atomic structure and use the same R-matrix method a better agreement for transitions among a larger number of levels is expected. 

The comparisons of $\Upsilon$ shown in Figs. 4 and 5 are only for transitions among the lowest 100 levels, as the aim is to provide a clear idea of the discrepancies. Considering all 28~203 transitions among the 238 levels, about 41\%, 38\% and 44\% of these differ by over 20\% at TE1, TE2 and TE3, respectively. Furthermore, not only are the values of $\Upsilon_{ICFT}$ larger in a majority of cases, the discrepancies are also greater than shown in Figs. 4 and 5, namely up to four orders of magnitude. Examples of such transitions are 59 -- 79/85/86 (2p3d~$^1$P$^o_1$ -- 2s6s~$^3$S$_1$, 2s6d~$^3$D$_1$, 2s6d~$^3$D$_2$) and there is no ambiguity in the ordering of these levels in our calculations and those of \cite{icft}. All of these (and many other) are inter-combination transitions, but the A- values between the two sets of calculations agree within 20\%.  Furthermore, the f-values for these  are $\sim$10$^{-6}$ and therefore, such weak transitions should behave as forbidden. Indeed this is the case in our calculations but not in those of \cite{icft}. Additionally, since  one of the authors (Luis Fern{\'a}ndez-Menchero) has kindly provided the $\Omega$ data for these transitions, we can confirm that  the differences in $\Upsilon$ values are not due to resonances.  Nevertheless, their $\Upsilon$ at TE3 are 2.4$\times$10$^{-1}$, 1.4$\times$10$^{-1}$ and 2.3$\times$10$^{-1}$, respectively, compared to our results of 3.1$\times$10$^{-3}$, 2.5$\times$10$^{-3}$ and 4.1$\times$10$^{-3}$, respectively. Similar discrepancies are found towards the lower end of the temperature range, and are partly due to  different $\Omega_B$, but mostly due to incorrect trends.

Finally, we discuss just one more example. For the 30--232/233/235 (2s4s~$^1$S$_0$ -- 2p7d $^3$D$^o_3$, $^3$P$^o_2$, $^3$P$^o_0$) transitions, the values of $\Upsilon_{ICFT}$ are larger than $\Upsilon_{DARC2}$ by about two orders of magnitude. These transitions are {\em forbidden} and resonances have (practically) zero  contribution. Therefore, both $\Omega$ and $\Upsilon$ should decrease with increasing energy/temperature. This  is the case in our work, but not that of \cite{icft}. Between T$_e$ =  3.2$\times$10$^3$ and 1.6$\times$10$^6$~K, the $\Upsilon$ of \cite{icft} increase from 1.08$\times$10$^{-3}$, 8.38$\times$10$^{-4}$, and 1.77$\times$10$^{-4}$ to 9.43$\times$10$^{-3}$, 1.07$\times$10$^{-2}$ and 2.18$\times$10$^{-3}$, respectively, whereas our results decrease from 5.24$\times$10$^{-4}$, 5.83$\times$10$^{-4}$, and 1.13$\times$10$^{-4}$ to 7.99$\times$10$^{-5}$, 1.13$\times$10$^{-4}$ and 2.11$\times$10$^{-5}$, respectively. Unfortunately, for these transitions also neither the $\Omega_B$ nor the trends in the ICFT calculations are correct.  Therefore, based on the comparisons shown here in Figs. 4 and 5, and the above discussion as well as those in \cite{belike, c3}, we confidently believe that the $\Upsilon$ results listed by \cite{icft} are indeed overestimated for a large number of transitions and over the entire range of temperatures. The reasons for this could be several as already stated \citep{belike, c3}. To recapitulate, these may be: (i) using two different ranges of partial waves with differing amount of $\Delta$E in the thresholds region, (ii) extrapolation of $\Omega$ over a very large energy range, and/or (iii) presence of some very large spurious resonances. In particular, we stress that the errors may be in the implementation of the ICFT, not the methodology itself.  Indeed this is (in a way) confirmed by the $\Omega$ data provided by Luis Fern{\'a}ndez-Menchero, because (i) is unlikely to be a major source of error and the authors of the ICFT calculations have checked for (iii), which also does not apply particularly to the 30--232/233/235 transitions. Apart from the high energy behaviour of $\Omega$ in the ICFT calculations, their approach (of converting the $LS$ results into $LSJ$) unreasonably affects the background values of  $\Omega$ for some inter-combination transitions.

\section{Conclusions}
In this work we have performed two sets of calculations  for energy levels, radiative rates, collision strengths, and most importantly effective collision strengths (equivalently electron impact excitation rates) for transitions in Be-like N~IV. In the first model, 166 levels of the $n \le$ 5 configurations are considered, whereas the second one is larger with 238 levels,  up to $n$ = 7. This is mainly to assess the impact of a larger model over that of a smaller one on (particularly) the determination of $\Upsilon$  and  to make a direct comparison with the similar calculations of \cite{icft}.

For the determination of energy levels and A-values,  the {\sc grasp} code has been adopted, and (the standard and parallelised versions of) {\sc darc} for the scattering calculations. These calculations are similar in methodology to our earlier work on other Be-like ions \citep{belike, c3},  but much larger. For the lowest 10 levels, discrepancies in energies with  measurements are up to 6\%, but agreement is better than 1\% for the remaining 228. Additionally, there are no significant discrepancies, in both magnitude or orderings, between our work and that of \cite{icft}. The A-values for E1, E2, M1 and M2 transitions have also been reported. For most transitions there are no (major) discrepancies between the two models or with other available data,  particularly for a majority of the strong E1 transitions. Lifetimes calculated with these A-values are also found to be in good agreement  with other available theoretical and experimental work, and hence (to an extent) confirm the accuracy of our calculations. Based on several comparisons as well as the ratio of the velocity and length forms of the A-values, our listed results are probably accurate to better than 20\% for a majority of the strong E1 transitions. 

Data have also been reported for collision strengths  over a wide range of energy, but only for resonance transitions. However, corresponding results for effective collision strengths are listed for all transitions among the 238 levels of N~IV and over a wide range of temperature up to 2.0$\times$10$^{6}$ K, well in excess of what should be needed for  modelling  astrophysical and fusion plasmas. In  our smaller model (DARC1) the energy resolution for resonances in thresholds region is comparatively coarser (0.001~Ryd), but is very fine in the larger one, i.e.  0.000~045~Ryd. Nevertheless,  for most transitions among the lowest 78 levels, there are no major discrepancies between the two sets of $\Upsilon$. However, discrepancies with the corresponding results of \cite{icft} are very large (up to four orders of magnitude) for over 40\% of the transitions, and over the entire temperature range. These discrepancies are similar to those already found for other Be-like ions \citep{belike, c3}, and do not support the conclusion of \cite{al10} that these are due to the size of a calculation. Our assessment is that for a majority of transitions, particularly among most of the lower levels, a larger calculation may improve the accuracy of $\Upsilon$, but the differences should not be very large. Therefore, the discrepancies found for transitions in many Be-like ions are not due to differences in the size of the atomic structure, but rather  the implementation of the method for calculating data. Based on several comparisons shown here and in previous papers,  we confidently believe that for most transitions the $\Upsilon$ data of \cite{icft} for Be-like ions are much overestimated. As this is perhaps most likely due to the implementation of ICFT, rather than the code itself,  a re-examination of their calculations would therefore be helpful.  

\section*{Acknowledgments}

This work has been carried out within the framework of the EURO fusion Consortium and has received funding from the Euratom research and training programme 2014--2018 under grant agreement No 633053 and from the RCUK Energy Programme (grant number EP/I501045). The views and opinions expressed herein do not necessarily reflect those of the European Commission. We are very thankful to our colleague  Dr. Connor Ballance for his help in generating data from the DARC2 calculations.  We also thank Dr. Luis Fern{\'a}ndez-Menchero for his kindness in providing $\Omega$ data for a few transitions and some clarifications about the ICFT calculations.

\vspace*{-0.3 cm}

\end{document}